\documentclass[multphys,vecphys]{svmult}

\usepackage{makeidx}     % allows index generation
\usepackage{graphicx}    % standard LaTeX graphics tool
\usepackage{multicol}    % used for the two-column index
\makeindex             % used for the subject index
\def\gtsim{\lower.5ex\hbox{$\buSildrel > \over\sim$}}
\def\ltsim{\lower.5ex\hbox{$\buildrel < \over\sim$}}
\def\arcsec{^{\prime\prime}}

\def\deg{{$^\circ$}}
\def\sun{\mbox{$_\odot$}}
\def\apjl{ApJL}
\def\apj{ApJ}
\def\apjs{ApJS}
\def\mnras{MNRAS}
\def\araa{ARAA}
\def\aj{AJ}
\def\aap{A\&A}
\def\aaps{A\&A Suppl.}
\def\nat{Nature}
\def\pasp{PASP}

\begin{document}

\title*{The Fueling and Evolution of AGN: Internal and External Triggers}
%\title*{Internal and External Triggers of AGN Activity}
%Internal and External Triggers of \\
%AGN and Starburst activity}
\titlerunning{Fueling AGN and Starbursts}
\author{Shardha Jogee\inst{1}
%\and Name of Author\inst{2}
}
% \authorrunningAGN: Internal and External Triggers}
\institute{Space Telescope Science Institute, 3700 San Martin Drive, 
Baltimore MD 21218, U.S.A 
\texttt{jogee@stsci.edu}
%\and Name and Address of your Institute \texttt{name@email.address}
}
\maketitle

\section{Introduction}

The  quest for a coherent picture of nuclear activity 
has witnessed giant leaps in the last decades.  
Four decades  ago, the idea was put forward
that accretion of matter  onto a  massive compact object 
or a supermassive black hole (SMBH) of mass 
$>$ 10$^6$  M$_{\odot}$ could  power  
very luminous active galactic nuclei (AGN), in particular, 
quasi-stellar objects (QSOs) (Lynden-Bell 1969; Soltan 1982; Rees 1984).
In the last decade,  dynamical evidence increasingly suggests 
that  SMBH pervade the centers of most massive galaxies ($\S$ 2 and
references therein). The challenge   has now shifted towards probing 
the fueling  and evolution of AGN 
over a wide range of cosmic lookback times,
and elucidating how they relate to their host galaxies in both
the local and cosmological context.

In this review, I will focus on the fueling and evolution 
of AGN under the influence of internal and external triggers.
In the nature versus nurture paradigm,  I use the term internal 
triggers to refer to intrinsic properties of host galaxies 
(e.g., morphological or Hubble type, color, and non-axisymmetric features
such as large-scale bars and  nuclear bars) 
while external triggers refer to  factors such as environment 
and interactions. The distinction is an over-simplification as
many of the so called intrinsic properties of galaxies can be 
induced or dissolved under the influence of external triggers.
Connections will be explored between the  nuclear and 
larger-scale properties of AGN, both locally and at intermediate 
redshifts.  One of the driving objectives is  to understand 
why not all relatively massive  galaxies  show signs of AGN  
activity (via high-excitation optical lines or X-ray emission) 
despite mounting dynamical evidence  that they harbor SMBHs. 
The most daunting challenge in fueling AGN  is arguably the 
angular momentum problem ($\S$ 3.2). Even  matter located  at a radius 
of a  few hundred pc must lose more than 99.99\%  of its 
specific angular momentum before it is fit for  consumption 
by a BH. 
%Thus, while there may be ample material in the inner few 
%hundred pc  to fuel a typical average Seyfert at 10$^{-2}$ M$_{\odot}$  
%yr$^{-1}$ over nominal  duty cycles,  the real  challenge is 
%the  angular momentum barrier rather than the mass of fuel.

The sequence of this review is as follows.  
$\S$~2 briefly addresses 
BH demographics and the  BH-bulge-halo correlations. 
$\S$  3 sets the stage for the rest of this paper by 
providing an  overview of  central issues in  the 
fueling of AGN and  circumnuclear starbursts.
In particular,  I review  mass accretion rates,  angular momentum
requirements, the effectiveness of different fueling mechanisms, and 
the growth and  mass density of BHs at different epochs.
% such as  gravitational torques,  
%(e.g., from large-scale and nuclear bars), 
% dynamical friction,  
%(on an accreting satellite  or on massive circumnuclear gas clumps),  
% hydrodynamical torques/shocks, and viscous torques. 
% The starburst--AGN connection is briefly touched. 
These central issues in $\S$  3 are attacked in more detail in 
 $\S$ 4--9  which describe different fueling mechanisms including
 mergers and interactions ($\S$ 5), 
large-scale bars ($\S$ 6), 
nuclear bars  ($\S$ 7),  nuclear spirals ($\S$ 8), and 
processes relevant on  hundred pc to sub-pc scales  ($\S$ 9).
I conclude with a summary and future perspectives in $\S$ 10. 
Complementary reviews  on  mass transfer and central activity  
in galaxies include those by Shlosman (2003), Combes (2003),
Knapen (2004), and Martini (2004).

\section{BH Demographics and BH-Bulge-Halo Correlations}

\subsection{Measurement of BH Masses}

The term SMBHs  refers to BHs having  masses  
  $M_{\rm bh}$  $>$ 10$^6$  M$_{\odot}$,  in contrast to  
intermediate mass BHs (IMBHs) with  $M_{\rm bh}$ $\sim$ 10$^2$--10$^6$  
M$_{\odot}$, and  stellar mass BHs. 
Properties of SMBHs are generally studied through accretion
signatures of  BHs or their gravitational influence.
The strongest dynamical evidence for SMBHs are in our Galaxy and 
in NGC~4258.  In these systems,  the large central densities 
inferred within a small resolved radius can be accounted for
by a SMBH,  but  not by  other possibilities such as 
collections of compact objects, star clusters, or exotic particles.
In  our Galaxy,  proper motion measurements set stringent constraints on  
the central potential (Sch{\" o}del et al. 2003; 
Ghez et al. 2003; Genzel et al. 
2000), yielding  $M_{\rm bh}$~$\sim$~3--4 $\times$ 10$^6$ M$_{\odot}$. 
% within 
%about  1000 $R_{\rm s-bh}$.  
In NGC 5248, VLBA  maser observations reveal Keplerian motions 
implying $M_{\rm bh}$~$\sim$~$3.9 \times 10^7$  M$_{\odot}$  (Miyoshi et 
al. 95).

In the last decade, high resolution gas and stellar dynamical 
measurements  from  ground-based
(e.g., Kormendy  \& Richstone 1995) and $HST$   observations  
(e.g., Harms et al. 1994; 
Ferrarese et al. 1996 ; van de Marel \& van den Bosch 1998; 
Ferrarese \& Ford 99; Gebhardt et al. 2000) 
have provided compelling evidence that several tens of galaxies 
host massive central  dark objects (CDOs) which are likely to be 
SMBHs.   
The more reliable dynamical measurements tend to be from observations 
which resolve the  radius of influence ($R_{\rm g-bh}$) 
within which the gravitational force of the BH exceeds that 
of nearby stars with  velocity dispersion $\sigma$, namely, 
\begin{equation}
 R_{\rm g-bh} = \frac{G M_{\rm bh}}{\sigma^{2}}
 = 11.2  \ {\rm pc} 
   \left  (\frac{M_{\rm bh}} {{ \rm 10^{8} \  M_{\odot} } }\right) 
   \left  (\frac{\sigma} {\rm 200 \  km \ s^{-1}}\right)^{-2}
\end{equation}
 
However, the scales probed by these measurements are still 
several 10$^5$--10$^6$ 
times the  Schwarzschild radius ($R_{\rm s-bh}$) of the BH, namely, 

\begin{equation}
R_{\rm s-bh}= \frac{2 G M_{\rm bh}}{c^{2}}
 = 5 \times 10^{-4} \ {\rm pc} 
   \left  (\frac{M_{\rm bh}} {{ \rm 10^{8} \  M_{\odot} } }\right) 
\end{equation}

The majority of the afore mentioned reliable measurements target  
ellipticals and a few  
early-type (Sa-Sbc) spirals  with central $\sigma$   $<$ 60  km~s$^{-1}$, 
and probe BH masses in the range  10$^7$--10$^9$  M$_{\odot}$.
Conversely, measuring BH masses in late-type spirals and dwarf galaxies 
poses many challenges, and  there are no firm measurements of BH masses 
below 10$^6$  M$_{\odot}$.
However, theoretical models and a mounting body of observational
evidence put the existence of IMBHs on a relatively firm 
footing  (see review by van der Marel 2003).
The first challenge in measuring the masses of IMBHs 
is that the gravitational radii of such BHs are typically too small 
to be easily resolved even with $HST$. 
A second complication is that  late-type spirals and dwarf galaxies 
which might harbor such BHs  also tend to host a bright 10$^6$--10$^7$  
M$_{\odot}$  stellar cluster  (Boker et al. 1999) whose dynamical effect 
can mask that of the  BH. 
A  10$^4$--10$^5$  M$_{\odot}$ BH (Filipenko \& Ho 2003) 
has been invoked  in the Sm dwarf NGC~4395 which hosts the 
nearest and lowest luminosity   Seyfert 1 nucleus. 
% (Filipenko \& Sargent 1989). 
Upper limits on BH masses are reported in several systems, e.g., 
10$^6$--10$^7$  M$_{\odot}$  for  six dwarf ellipticals in 
Virgo (Geha, Guhathakurta, \& van der Marel 2002), 
$5 \times 10^5$  M$_{\odot}$  for the Scd spiral IC342 (Boker et al.1999).
Gebhardt, Rich, \& Ho (2002)  infer the presence of an IMBH with a mass  
of a  few $\times$ 10$^4$  M$_{\odot}$  in  one of the most massive 
stellar clusters (G1) in M31, but an alternative interpretation 
of  the dataset has been presented by Baumgardt et al. (2003). 
A tantalizing dark central mass concentration of 
a few $\times 10^3$ M$_{\odot}$  (Gerssen et al. 2003) 
is reported in the globular cluster M15 from $HST$ data,  but it 
remains unclear whether it is an IMBH. 
$Chandra$ observations of ultraluminous X-ray sources also 
suggest the presence of IMBHs (Clobert \& Miller 2004 and
references therein).

At many levels, measuring BH masses in local AGN  such as 
Seyferts and LINERS is more challenging than corresponding 
measurements in massive quiescent galaxies. 
The bright non-thermal active nucleus in Seyfert galaxies 
can drown  the spectroscopic features from which dynamical 
measurements are made. Consequently, BH masses in local AGN 
are commonly  mapped with  alternative techniques such
as reverberation mapping (Blandford \& Mc Kee 1982; 
Peterson 1993; see Peterson these proceedings) where one 
estimates the virial mass  inside the broad-line region (BLR)
by combining the velocity of the BLR 
with an estimate of the size of the BLR based on time
delay measurements. Reverberation mapping  can typically 
probe scales $\sim$ 600 $R_{\rm s-bh}$ 
and has yielded  BH masses
for several tens of AGN   (Peterson 1993; Wandel, Peterson, \& Malkan 
1999; Kaspi 2000). Earlier controversies existed on the 
reliability of the method due to   purported 
systematic differences in the  BH-to-bulge mass ratio  
between  AGN  with reverberation mapping data  and 
quiescent galaxies or QSOs. 
However, recent work (e.g., Ferrarese et al.  2001)  claims 
that  for AGN with accurate 
measurements of stellar  velocity dispersions, 
the reverberation 
masses agree with the BH mass determined from the tight
$M_{\rm bh}$--$\sigma$ relation ($\S$ 2.2) which is derived 
from quiescent galaxies. 
% It now appears that the ratio of BH to bulge masses 
% ($M_{\rm bh}$/$M_{\rm bulge}$) in quiescent galaxies, 
% AGNs, and QSOs all appear to 
% be consistent at 0.13 \%, 0.09%, and 0.12\% respectively

\subsection{Relationship of the Central BH to the Bulge and Dark Halo}
% {BH-Bulge-Dark Matter Halo correlation}

\begin{figure}
\centering
%\includegraphics[width=11cm]{geb.fg2.h.eps}
\includegraphics[width=11cm]{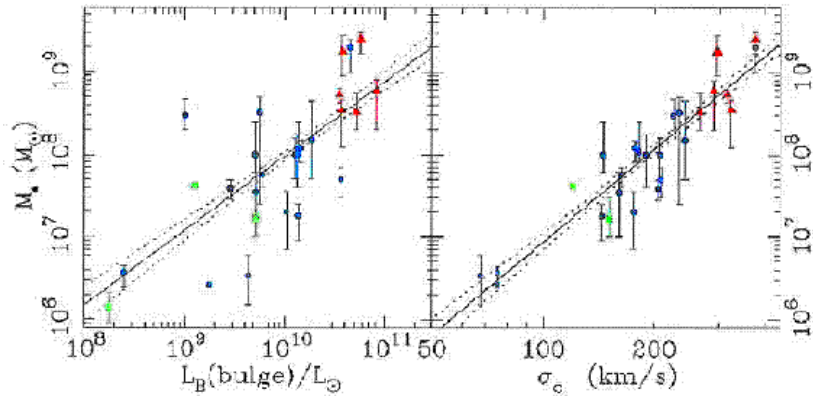}
%\picplace{5cm}{2cm} % Give the correct figure height and width in cm
%
\caption{ 
\bf Correlation between central BH mass and circumnuclear 
velocity dispersion  --  \rm
Black hole mass versus bulge luminosity (left) and the luminosity-weighted aperture dispersion within the effective radius (right). 
Green squares denote galaxies with maser detections, red triangles are 
from gas kinematics, and blue circles are from stellar kinematics. 
Solid and dotted lines are the best-fit correlations and their 
68 \% confidence bands. (From Gebhardt et al. 2000)
}
\label{fig:1}       % Give a unique label
\end{figure}

A tight correlation has been  reported between the mass of 
a central BH and the stellar velocity dispersion ($\sigma$) of the host 
galaxy's  bulge (Ferrarese \& Merritt 2000; Gebhardt et al. 2000): 

\begin{equation}
M_{\rm h} = \alpha  
\left(\frac {\sigma} { {\rm 200 \ km \ s^{-1} }}\right)^{\beta}
\  {\rm M_{\odot}} 
\end{equation}

where 
$\alpha$ = (1.7 $\pm$ 0.3) $\times 10^8$,  $\beta$ =(4.8 $\pm$ 0.5)
(Ferrarese \& Merritt 2000), and   
$\alpha$ = (1.2 $\pm$  0.2) $\times 10^8$, $\beta$ = (3.8 $\pm$ 0.3) 
(Gebhardt et al. 2000). 
Tremaine et al. (2002) assign the range in quoted values for
$\beta$  to systematic differences in velocity dispersions 
used by different groups. 
The   $M_{\rm bh}$--$\sigma$ relation reported  originally 
in the literature  
(Gebhardt et al. 2000; Ferrarese \& Merritt 2000; Tremaine et al. 2002)
is primarily based on local early-type galaxies  (E/SOs) and a  
handful  of spirals Sb--Sbc, and it primarily samples  quiescent 
BHs with  masses in the range  a few $\times$ (10$^7$--10$^9$) 
M$_{\odot}$.  This relation was subsequently found to also hold in  
AGN hosts  (Ferrarese et al.  2001), and in 
bright QSOs out to $z$~$\sim$~3  with estimated  BH masses of up 
to 10$^{10}$  M$_{\odot}$  (Shields et al. 2003).
\it
This suggests that active and quiescent BHs  bear a common  
relationship to  the surrounding triaxial component of their host 
galaxies over a wide  range of  cosmic epochs and BH 
masses \rm (10$^6$--10$^{10}$ M$_{\odot}$). 
\rm

Numerous variants of the $M_{\rm bh}$--$\sigma$  relation 
have been proposed. 
While earlier correlations between 
the  mass of CDOs/SMBHs and the bulge luminosity ($L_{\rm bulge}$) 
had significant scatter (e.g., Kormendy \& Richstone 1995), 
recent work (H{\" a}ring \& Rix 2004)  
based on improved  BH and  bulge masses  yield a very tight 
$M_{\rm bh}$--$M_{\rm bulge}$ relation.
% or  between the BH and bulge mass in AGN  (Wandel 1999).  
Graham et al. (2001) find a correlation 
between the light concentration of galaxies and the mass of 
their SMBHs,  and claim this relation is as tight as 
the $M_{\rm bh}$--$\sigma$  relation. 
Grogin et al. (2004) have searched for signs of this correlation at
$z\sim$~0.4--1.3 in    a comparative study of structural parameters 
among 34000 galaxies in the GOODS fields, including 350 X-ray selected AGN 
hosts in the overlapping  $Chandra$ Deep Fields.
Compared to the inactive galaxies, the AGN hosts have significantly 
enhanced concentration indices throughout the entire redshift range, 
as measured in rest frame $B$-band for a volume-limited sample 
to $M_{\rm B}$ $<$ -19.5 (and to $L$(2--8keV) $> 10^{42}$  for the
AGN).
Finally, Ferrarese (2002) shows that the 
$M_{\rm bh}$--$\sigma$  relation translates to a 
relation between the mass of the BH  and that of the 
dark matter (DM) halo ($M_{\rm dm}$)   

\begin{equation}
M_{\rm h} \  =   10^{7} \ {\rm M_{\odot}} 
\left(\frac{M_{\rm dm}} { 10^{12}\ {\rm M_{\odot}} }\right)^{1.65}
\end{equation}

if one assumes that $\sigma$  correlates with the circular 
speed $V_{\rm c}$ which  bears an intimate relation  to 
the DM  halo within the standard  $\Lambda$CDM paradigm.

A plethora of theoretical studies have explored the growth of 
BHs and the possible origin of a fundamental $M_{\rm bh}$--$\sigma$ relation
(e.g., Haehnelt \& Kauffmann 2000; Adams, Graff, \& Richstone 2001; 
Burkert  \& Silk 2001; Di Matteo, Croft, Springel, \& Hernquist 2003; 
Bromm \& Loeb 2003; Wyithe \& Loeb 2003; El-Zant et al. 2003).
According to  Haehnelt \& Kauffmann (2000), hierarchical galaxy 
formation models where  bulges and SMBHs both form during major mergers  
produce a $M_{\rm bh}$--$\sigma$ correlation.
Star-formation (SF)  regulated growth of BHs in protogalactic spheroids 
has been proposed by Burkert  \& Silk (2001) and  Di Matteo et al. (2003). 
In many of these models, black hole growth stops because of the
 competition with SF  and, in particular, 
feedback, both of which determine the gas fraction available for accretion. 
According to Wyithe \& Loeb (2003),  a  tight 
$M_{\rm bh}$--$\sigma$  relation  naturally results  from 
hierarchical  $\Lambda$CDM   merging  models  where  
SMBHs in galaxy centers 
undergo  self-regulated  growth within  galaxy halos until 
they unbind the galactic gas that feeds them.  
% and the characteristic quasar 
% lifetime of ~10$^{7}$ years simply corresponds to  the dynamical time 
% of galactic disks during the epoch of peak quasar activity ($z \sim$2.5).
% Since the lifetime becomes comparable to the Salpeter e-folding time 
% at this epoch, the model also implies that the Mbh-sigma relation is a 
% product of feedback-regulated accretion during the peak \left of quasar 
% activity. 
El-Zant et al. (2003) have suggested that the BH--bulge--DM halo 
correlation  can be understood within the framework of galactic 
structures growing within flat-core, mildly triaxial halos.

\section{Central Issues in Fueling AGN and Starbursts} 

I present here an overview of several central issues that 
are relevant for understanding the  fueling of AGN and  
circumnuclear starbursts.  

\subsection{Mass Accretion Rates}

For a standard BH accretion disk  with  an  efficiency
$\epsilon$ of conversion between matter and energy, the 
radiated  bolometric luminosity  $L_{\rm bol}$ is related to 
the mass accretion rate ($\dot{M_{\rm bh}}$) at the last stable 
orbit of a  BH by   

\begin{equation}
  \dot{M_{\rm bh}}
 =  0.15 \ {\rm M_{\odot} \ yr^{-1}} 
    \ \left(\frac  {0.1} { \epsilon } \right)
    \ \left(\frac { L_{\rm bol} } { {\rm 10^{45} \ ergs \ s^{-1}}}\right)
\end{equation}
% (e.g., Shapiro \& Teukolsky 1983). 

Table 1 shows typical observed bolometric luminosities and  inferred 
mass accretion rates  
for QSOs and local AGN  (Seyfert, LINERS) 
assuming  a standard radiative efficiency $\epsilon$~$\sim$ 0.1.
The standard value of  $\epsilon \sim$ 0.1   applies 
if  the    gravitational binding energy liberated 
 by the accreting gas  at the last stable orbit of the BH 
 is radiated  with an 
efficiency of  $\sim$  0.1 $c^{2}$.
%  The last stable circular orbit where gravitational energy 
%  is released  has a radius which is related to the 
%  Schwarzchild radius ($R_{\rm sch}$) of the BH by a numerical factor 
%  $\alpha$  that depends on the geometry and  rotation of the accretion disk'
%   $\alpha$  = 3 for standard acc disk.
%  \begin{equation}
% R_{\rm sch}
%     =  2  \frac { G M } {c^{-2} }
% \end{equation}
In practice, the radiative   efficiency depends  on the 
nature of the  accretion disk and   gas accretion  
flows. 
For instance,  thin-disk accretion onto a Kerr BH can 
lead to a  radiative   efficiency $\epsilon$~$\sim$~0.2. 
It has been suggested that the most luminous quasars at high 
redshift   may have grown  with  $\epsilon$~$\sim$~0.2,   or 
alternatively that they have a super-Eddington luminosity 
(Yu \& Tremaine 2002). 
Conversely, in certain popular models of gas accretion 
flows such as adiabatic inflow-outflow solutions 
(ADIOS; Blandford \& Begelman 1999) and   convection-dominated 
accretion flows (CDAF: Narayan et al. 2000) 
%which 
%are variants on advection-dominated  accretion flows 
% (ADAF; Narayan, Mahadevan \& Quataert  1998), 
only a small fraction of the matter which accretes
at the outer boundary  of the flow   contributes 
to the mass accretion rate at the BH  
due to turbulence and strong mass loss.
This leads to an effective radiation 
efficiency  $\ll$ 0.1   when applied to  the mass accretion 
rate %($\dot{M_{\rm outer}}$)
at the  outer boundary of the accretion flow.  
Thus, within the  CDAF and ADIOS 
paradigms,  the gas inflow rates that must be supplied on 
scales of tens of pc may be much larger than those quoted in 
Table 1, even for low luminosity Seyferts.

\begin{table}
\centering
\caption{Typical $L_{\rm bol}$ and $\dot{M_{\rm bh}}$ for QSOs and local AGN}
\label{tab:2}       % Give a unique label
\begin{tabular}{cccc}
\hline\noalign{\smallskip}
Type of AGN & 
~~~~~~$L_{\rm bol}$$^{\rm a}$ & 
~~~~~~Typical $L_{\rm bol}$ &
Typical $\dot{M_{\rm bh}}$$^{\rm b}$ \\
&
~~~~~~(ergs s$^{-1}$) & 
~~~~~~(ergs s$^{-1}$) & 
(M$_{\odot}$ yr$^{-1}$) \\
(1)  & 
~~~~~~(2)  &  
~~~~~~(3) &
(4) \\
\noalign{\smallskip}\hline\noalign{\smallskip}
QSOs & 
~~~10$^{46}$--10$^{48}$ & 
~~~~~~~~10$^{47}$--10$^{48}$ &
10--100  \\ 
Seyferts & 
~~~10$^{40}$--10$^{45}$   & 
~~~~~~~10$^{43}$-10$^{44}$ &
~10$^{-3}$--10$^{-2}$ \\
LINERs & 
~~~~10$^{39}$--10$^{43.5}$   & 
~~~~~~~~10$^{41}$--10$^{42}$  &
~10$^{-5}$--10$^{-4}$ 
\\
\noalign{\smallskip}\hline
\multicolumn{4}{l} 
{\small 
Notes to Table -- 
a.  The full range in  bolometric luminosity ($L_{\rm bol}$) 
for Seyfert 
} \\
\multicolumn{4}{l} 
{\small 
and LINERS is taken  from  Ho, Filippenko, \& Sargent 1997a, while 
for QSOs
} \\
\multicolumn{4}{l} 
{\small 
different sources in the literature are used; b. The typical 
$\dot{M_{\rm bh}}$ in column (4) 
} \\
\multicolumn{4}{l} 
{\small 
is derived from the typical $L_{\rm bol}$ in column (3) assuming a  
standard  radiative 
} \\
\multicolumn{4}{l} 
{\small 
efficiency $\epsilon$~$\sim$~0.1
}\\
\end{tabular}
\end{table}

\subsection{The Angular Momentum Problem}

\bf \it
The most important challenge in fueling AGN is 
the angular momentum problem
\rm
rather than the amount of fuel \it per se. \rm
The angular momentum per unit mass  or specific 
angular momentum $L$=$r \times v $ of fuel 
at the last stable radius of 
a BH of mass ($M_{\rm 8} \times 10^8$ M$_{\odot}$)
is several times $10^{24}  M_{\rm 8}$  cm$^{2}$~s$^{-1}$.
In contrast, matter (star or  gas)  rotating in  
a spiral or elliptical galaxy at a radius of 10 kpc has 
a specific angular momentum of several times 
$10^{29}  M_{\rm 8}$ cm$^{2}$~s$^{-1}$. 
This is illustrated  in Fig. 2  assuming 
typical galactic  rotation velocities. 
Thus,
\it  
the specific angular momentum of  matter located at a radius of a 
few kpc must be  reduced by   more than 10$^{4}$   before 
it is fit for  consumption by a BH.
\rm 
Searching for mechanisms which can achieve this miraculous 
reduction of angular momentum is one of the driving objectives 
of  AGN research. 
Even at a radius of 200 pc, $L$  is still a factor of 1000 
too large,  and  the angular momentum barrier is a more 
daunting challenge than  the amount  of gas.
For instance, in the case of  a Seyfert  with  an accretion rate  of $\sim$  
10$^{-2}$ M$_{\odot}$ yr$^{-1}$  and a duty cycle of  
10$^{8}$ years,  a gas cloud  of 10$^{6}$ M$_{\odot}$  
may provide adequate fuel. Such clouds  are certainly common 
within the inner 200 pc  radius of spiral galaxies,
but we yet have to understand  what physical processes 
are able to  squeeze their  angular momentum out by 
more than 99.99\%. 
The BH  is analogous to an exigent dieter  who has a plentiful 
supply of rich food, but can only consume 99.9\% fat-free items!

\begin{figure}
\centering
\vspace{0.5 cm}
\hspace{-3.15 cm}
\includegraphics[height=17cm]{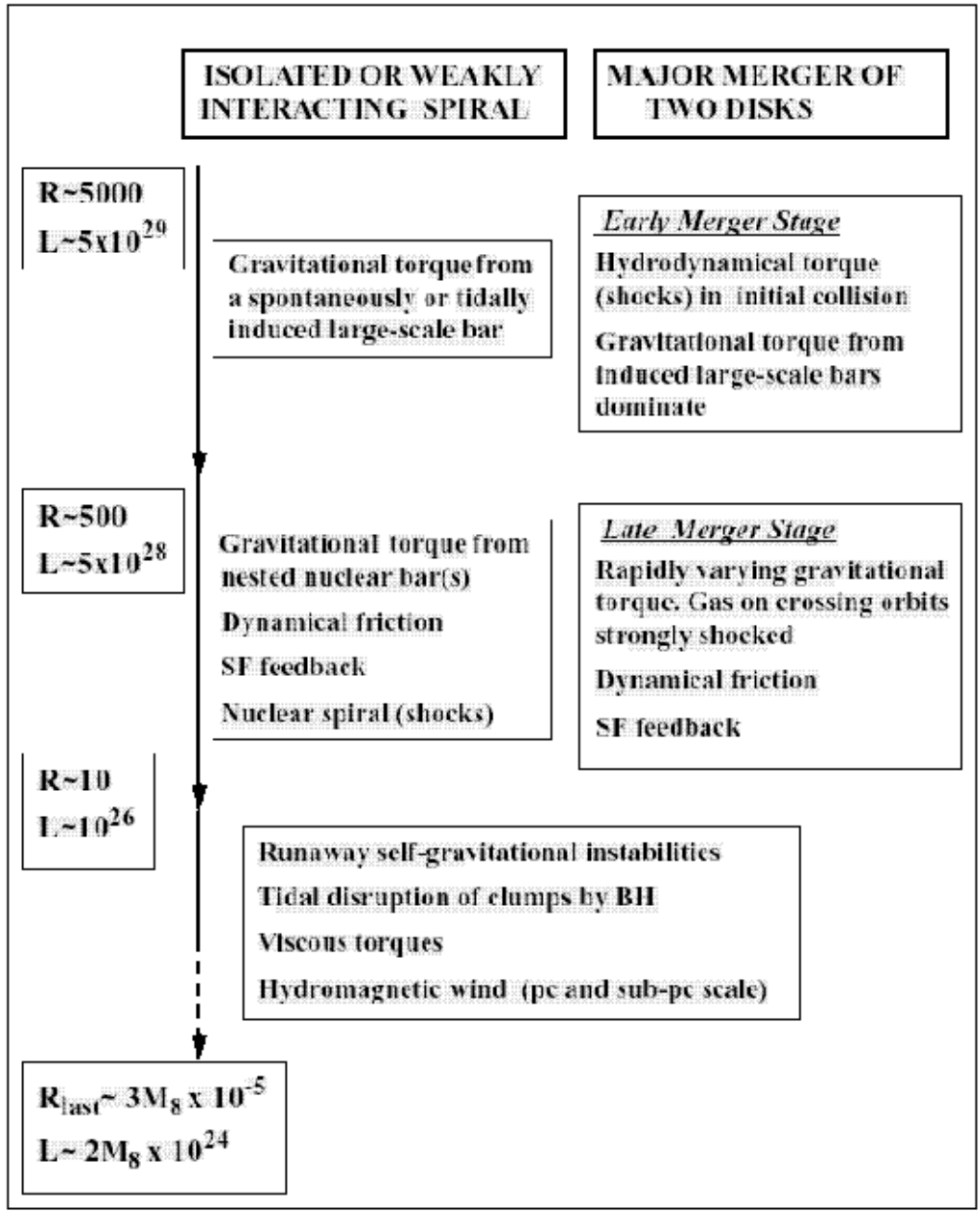}
\hspace{-2.15 cm}
\caption{ 
\bf The angular momentum problem in the fueling of AGN and 
starbursts: 
\rm
\rm
The specific angular momentum ($L$) of  gas located at a radius 
($R$) of several kpc must be  reduced by more than 10$^{4}$ before 
it is fit for  consumption at the last stable orbit ($R_{\rm last}$) 
of a BH.  In contrast, powerful starbursts can be  more easily 
triggered via gravitational torques which  build 
large gas densities on circumnuclear ($R$=500 pc)   scales.
This figure  schematically illustrates some  mechanisms that can  
reduce  $L$ and drive gas inflow on various spatial scales
in  a relatively quiescent  galaxy \it (left) \rm  and in a
major merger \it (right)\rm. $R$ is in pc, $L$ is in units 
of cm$^{2}$~s$^{-1}$,  and a ($M_{\rm 8} \times 10^8$ M$_{\odot}$)
BH is assumed. See text for details.
}
\label{fig:1}       % Give a unique label
\end{figure}

\subsection{Dominant Fueling Mechanisms on Different Scales}
\rm
~Gravitational torques, dynamical friction, viscous torques, and
hydrodynamical torques (shocks) are some of the mechanisms which 
remove angular momentum from the dissipative
gas component and channel it to small scales, thereby helping
to fuel central starbursts and massive BHs. 
These different  fueling mechanisms 
assume a different relative importance at different radii 
in a galaxy, and also, when dealing with a strongly interacting galaxy 
versus an  isolated one. 
I will review these different mechanisms in detail 
from an observational and theoretical perspective in  
$\S$ 4--9, but here I discuss a few key concepts  and 
provide a schematic overview in  Fig. 2.

\begin{table}
\centering
\caption{Gravitational Torques, Dynamical Friction, and Viscous Torques}
\label{tab:2}       % Give a unique label
\begin{tabular}{lllll}
\hline\noalign{\smallskip}
$R$ & 
$M$ & 
$t_{\rm gra}$  & 
$t_{\rm df}$ & 
$t_{\rm visc}$ \\
(pc)   &  
(M$_{\odot}$)  & 
(Myr) & 
(Myr) & 
(Myr) \\
(1)  & 
(2)  &  
(3) & 
(4) &
(5)\\
\noalign{\smallskip}\hline\noalign{\smallskip}
1000 & 
1e7  & 
20 & 
1020 & 
1000 \\ 
200 & 
1e7 & 
4 & 
60 &
- \\
\noalign{\smallskip}\hline
\end{tabular}
\end{table}

Gravitational torques  operate on  a timescale ($t_{\rm gra}$) 
comparable to the orbital timescale and  provide, therefore,  
the most efficient  way  of reducing  angular momentum 
on large to intermediate scales (tens of kpc -- a few 100 pc).
This can be seen by comparing $t_{\rm gra}$  with the typical 
timescales on which dynamical friction ($t_{\rm df}$) and 
viscous torques  ($t_{\rm vis}$) operate for a cloud of mass $M$ 
(Table 2). Dynamical friction  on a clump of mass 
$M$ and speed $v$  at a radius  $R$ operates on  a timescale
which is 
$\propto$ ($R^2$ $v/M$ ln$\Lambda$), 
where ln$\Lambda$ is the Coulomb logarithm (Binney \& Tremaine 1987). 
For a  $10^7$  M$_{\odot}$ gas cloud at a kpc radius  
in a disk galaxy, $t_{\rm df}$ is  an order of magnitude
larger than  $t_{\rm gra}$  (Table 2). 
However, for massive gas clumps  at low radii, dynamical friction 
becomes increasingly  important: it  can drive 
a 10$^{8}$  M$_{\odot}$  cloud from  $R\sim$~200 pc 
down to $R\sim$~10 pc 
within a few  times $10^7$ yrs  ($\S$ 9).

In an isolated  galaxy (Fig. 2), gravitational torques are exerted by 
non-axisymmetric  features such as  
large-scale  ($\S$ 6)  and nuclear  ($\S$ 7)  bars. 
While a large-scale bar efficiently drives gas from the outer disk 
into the inner kpc, the 
bar-driven gas flow slows or even stalls as it crosses the 
inner Lindblad  resonance (ILR) for reasons described in  $\S$ 6.  
At this stage,  the gas  piles up typically at a radius of   
several 100 pc where   powerful  starbursts  are commonly observed 
($\S$ 6; Fig 6).  
However, gas on these scales has a  specific angular momentum 
that is still more than 1000 times too high  for it to be 
digestible by a BH. 
If a  nuclear bar ($\S$ 7) is present, it can 
break the status quo and torque gas from the ILR region of the 
large-scale bar down to  tens of pc. 
In addition, if massive gas clumps  exist in the inner few 100 pc,  
dynamical friction can drive  them down to tens of pc ($\S$ 9).
Finally, feedback from SF  (e.g., shocks from 
supernovae)  can remove energy and angular momentum ($\S$ 9)
from a small fraction of the circumnuclear gas.
On  scales of tens of pc, the tidal torque from the BH itself 
can disrupt gas clumps and stellar clusters, possibly into an 
accretion disk ($\S$ 9). Subsequently,  on pc and sub-pc scales,  
viscous torques and hydromagnetic outflows in AGN  ($\S$ 9)
may become important.

Simulations suggest that  induced  large-scale stellar bars 
remain the main driver of gas inflows down to scales of a few 
100 pc, even in the case of interacting galaxies (Fig. 2), 
namely in many minor mergers ($\S$ 5.2) and during 
the \it early \rm stages of  major  (1:1) and intermediate mass ratio  
(1:3) interactions ($\S$ 5.1). 
Just like in the case of an isolated barred galaxy, 
gas inflows driven by an induced bar also slow down near 
the ILR.  However, the  \it final  stages \rm of a major or 
intermediate mass ratio  merger bring
in very different elements.  As violent relaxation starts, 
gas experiences  strongly-varying   gravitational torques, 
and if it is on interacting and crossing orbits, it also 
suffers strong  shocks  ($\S$ 5.1; Fig. 2). Thus, 
in the final  merger stages, gas loses angular momentum and  
large gas inflows ($\gg$ 1M$_{\odot}$ yr$^{-1}$) 
down to small scales  can result,   provided the earlier 
episodes of  SF  have not depleted most of the circumnuclear 
gas already ($\S$ 5.1).

\subsection{Census and Growth Epoch of BHs}

Table  3 compares the  BH mass density  ($\rho_{\rm bh-qso}$)   
accreted during the optically bright  QSO phases ($z$=0.2--5) 
to the  BH mass density in present-day galaxies (both active
and inactive). 
Yu \& Tremaine (2002)  find 
$\rho_{\rm bh-qso}$~$\sim$~(2.5 $\pm$ 0.4)~$\times$~10$^{5}$ 
($h_{\rm 0}$/65)$^{2}$ M$_{\odot}$ Mpc$^{-3}$
using the  extrapolated 
QSO luminosity function from the 2dF redshift survey 
and  a radiative efficiency of  0.1.
Similar values 
%in the range $2--4 \times$  10$^{5}$  M$_{\odot}$   Mpc$^{-3}$  
have been reported by others including 
Wyithe \& Loeb (2003),  Ferrarese (2002b), and 
Chokshi \& Turner (1992). 
This value  of $\rho_{\rm bh-qso}$ is a lower limit to 
the total BH mass density we expect to be in place by 
$z$=0.2  since it does not incorporate optically obscured QSOs  
and any build-up of the BH mass occurring outside the 
QSO phase.  However, it is probably not far off,  since 
the  BH mass density from X-ray AGN counts  at 
$z>0.2$  ($\rho_{\rm bh-xray}$) 
is  estimated to be  2--5 $\times$ 10$^{5}$  M$_{\odot}$ 
Mpc$^{-3}$ (Cowie \& Barger 2004; Fabian \& Iwasawa 1999; Table 3).

\begin{table}
\centering
\caption{Census of BH Mass density}
\label{tab:2}       % Give a unique label
\begin{tabular}{lcc}
\hline\noalign{\smallskip}
\multicolumn {3}{c} 
{BH~Mass~Density~[10$^{5}$ M$_{\odot}$  Mpc$^{-3}$]} \\
%(1)  & 
%(2)  \\
\noalign{\smallskip}\hline\noalign{\smallskip}
$\rho_{\rm bh-QSO}$ accreted during optical QSO phase  ($z$=0.2--5) &
~~~~~~~~2--4$^{a,b,c,d}$ \\  
$\rho_{\rm bh-Xray}$ from X-ray background ($z >0.2$) &
~~~~~~~~2--5$^{e,f}$ \\
% \noalign{\smallskip}\hline
$\rho_{\rm bh-local}$ in local early-type galaxies  ($z<0.1$) &
~~~~~~~~2--6$^{a,b,g}$ \\
$\rho_{\rm bh-Sy}$ in local Seyferts   &
~~~~~~~~$<$ 0.5$^{c}$
\\
\noalign{\smallskip}\hline
\multicolumn {3}{l} 
{\small 
References in table -- a. Yu \& Tremaine 2002; b. Wyithe \& Loeb 2003; 
} \\
\multicolumn {3}{l} 
{\small 
c. Ferrarese 2002; d. Chokshi \& Turner 1992; e. Cowie \& Barger 2004;
}\\
\multicolumn {3}{l} 
{\small 
f. Fabian \& Iwasawa 1999; 
g. Merritt \&   Ferrarese 2001.
}\\ 
\end{tabular}
\end{table}

In the local Universe, the BH mass density in early-type galaxies
at $z < 0.1$ is estimated   to be (2.5 $\pm$  0.4) $\times$ 10$^{5}$ 
($h_{\rm 0}$/65)$^{2}$ M$_{\odot}$   Mpc$^{-3}$, based on 
the measured  velocity dispersion  of early-type 
galaxies in the Sloan Digital Sky Survey 
and the  $M_{\rm bh}$--$\sigma$ relation (Yu \& Tremaine 2002).
% Similar results are reported by Wyithe \& Loeb (2003)
% and Merritt \&  Ferrarese  (2001).
However, rough estimates of the BH mass density in local 
\it active \rm Seyfert 1 and 2 
galaxies yield significantly lower values  
(Ferrarese 2002;  Padovani et al. 1990; Table 3).

In summary, the census of BH mass density (Table 3)  suggests 
that accretion with a standard radiation efficiency of 0.1  
during the quasar era can readily account for  the BH 
mass density found in  local ($z<0.1$)  early-type galaxies.
Only a small fraction of this  local BH mass 
density appears to be currently active as Seyfert galaxies
and the inferred mass accretion rates in such cases are typically 
 10$^{3}$ times lower than in QSOs.
This suggests that there
\it
 is no significant growth of BHs in 
the present epoch compared to the quasar era.
\rm
Thus, we should bear in mind that   local AGN (Seyferts)  
with current low levels of BH growth may well  differ  from 
luminous QSOs near $z\sim 2.5$ in one or more of the following 
characteristics: 
\it 
the nature 
of the dominant fueling mechanism, the amount of cold gas
reservoir, and the nature of the host galaxy.
\rm
For instance, tidal interactions and minor or major mergers  
may have been  much 
more important in the quasar era and early epochs of galaxy
growth than they are in activating present-day Seyfert galaxies.

\subsection{The Starburst--AGN Connection}
While I discuss  the fueling of both AGN and starbursts in 
this review, I will  not  explicitly address the  
starburst--AGN connection.
I only mention here that  this connection  can be 
\it 
circumstantial, influential, or  causal.
\rm 
A  \it circumstantial \rm 
connection refers to the fact that 
starburst and AGN activity can both manifest in the 
same system  simply because they are affected by a 
common element such as  a rich  supply of  gas, or an
external trigger (e.g., an interaction).
Examples include 
the ULIRG--QSO connection (Sanders et al. 1988), 
evolutionary  scenarios  for Seyfert 2 
(e.g.,  Storchi-Bergmann et al.  2001), and perhaps the blue 
color of AGN hosts described in $\S$ 4.
An \it influential \rm connection is one where 
the AGN and starbursts 
may contaminate  each other's   
\it observed \rm 
properties. 
Examples include the  starburst affecting  the featureless
continuum and line ratios of Seyferts (Cid-Fernandes et al. 2001),
or washing out the  hard accretion disk spectrum. 
A \it causal \rm 
connection  is a more fundamental connection  where the starburst 
causes the AGN or vice versa. One example is the evolution of 
a dense stellar cluster into a BH (Norman \& Scoville 1988).

\subsection{A Note of Caution on Empirical Correlations}

There exists many contradictory reports in the literature
of correlations or lack thereof between starburst/AGN 
activity and host galaxy properties (e.g., Hubble types, bar
fraction, nuclear bar fraction), or external triggers 
(e.g., presence of companions, morphological signs of 
interactions/mergers).  
Many caveats conspire towards this dismal state of affairs
and should be avoided: 
\it
(1)
\rm
Many early studies fail to  adopt the key practice of having
a large control sample which is matched to the active sample
or to the starburst sample in terms of relevant parameters 
such as distance, morphological types, luminosities, 
inclinations, and environments.
\it
(2)
\rm
The classification of  morphological features such
as bars and Hubble types is still often made  from optical catalogs 
(e.g., the Third Reference Catalogue (RC3); de Vaucouleurs et al. 
1991) and  suffer from subjectivity, low spatial resolution, and 
contamination by dust.  It is better to use  a quantitative 
method (e.g., ellipse fits) for characterizing bars and apply 
it to near-infrared (NIR) rather than optical images. The former are less
affected by extinction and typically yield a  bar fraction 
which is higher by 20--30 \% (e.g., Knapen et al. 
2000; Eskridge et al. 2002). 
\it
(3)
\rm
Cross comparisons of discrepant results are often difficult 
because they are based on inhomogeneous samples drawn from 
different local AGN catalogs that have limited overlap and 
different biases. 
For instance, optically selected magnitude-limited samples may 
be biased against 
faint nuclei embedded in bright galaxies. UV-based catalogs may 
favor blue Seyfert 1 and quasars. Commonly used  catalogs include 
the Veron-Cetty \& Veron  Catalog  of Seyferts and LINERS,
the optically selected CfA sample of 48 Seyferts (Huchra \& Burg 1992),
% 48 Sy = 27 type 1, 21 type 2
the Palomar Optical  Spectroscopic Survey  (POSS;  Ho  et al. 
1997a) of 486 emission line nuclei geared towards low 
luminosity H{\sc ii} and AGN nuclei, and 
the extended 12 $\mu$m  Galaxy Sample (E12GS) of 891 galaxies  
(Hunt \& Malkan 1999).
\it
(4)
\rm
Nuclear types (H{\sc ii}, LINER, Seyferts) listed in literature
databases such as NED often show significant discrepancies from 
recent careful spectroscopic classifications (e.g., Ho et al. 1997a).
In  $\S$ 4--9,  I will focus on studies which tend to avoid these
caveats  or alternatively qualify  the caveats as they arise.

\section{Hubble Type and Colors of AGN Hosts}

\begin{figure}
\centering
%\includegraphics[height=9cm]{hunt99.fg1.eps}
\includegraphics[height=9cm]{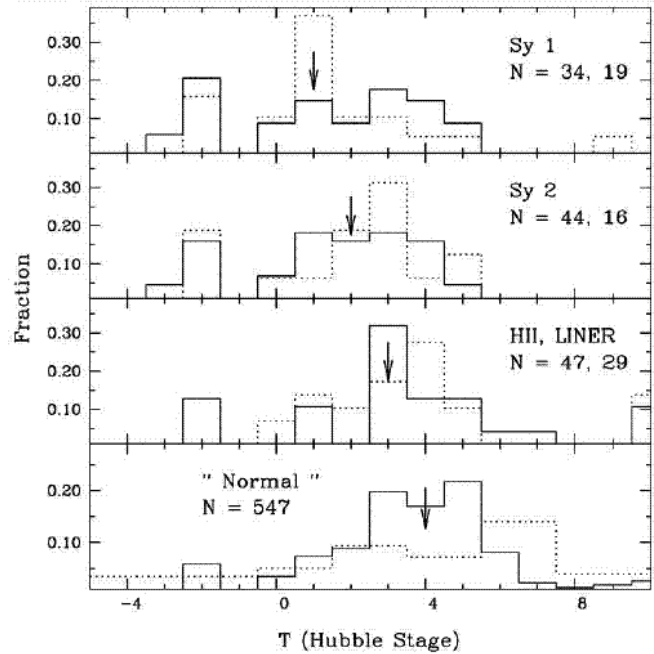}
\caption{ 
\bf Distributions of the  Hubble types in  AGN (Seyferts, LINERS),
    H{\sc ii} and normal nuclei: 
\rm 
The \it solid line \rm  in all four panels represents data from the 
12 $\mu$m sample (E12GS; Hunt \& Malkan 1999).  
The  \it dotted line \rm represents data
taken from  E12GS or other AGN catalogs. The panels 
represent \it (from top to bottom) \rm
(1) The Sy1 sample from E12GS  (solid line) and the CfA (dotted line) 
sample; (2) The Sy1 sample from E12GS  (solid) and the CfA (dotted) 
sample; (3) The  H{\sc ii}  (solid)  and LINER  (dotted)  
samples from E12GS; (4) The normal galaxies from  E12GS  (solid) 
and the Uppsala General Catalog  (dotted) 
as tabulated by Roberts \& Haynes (1994).
The numbers  in each panel  refer to the  number of objects  
represented by the solid and dotted histograms. 
The data have been binned in terms of RC3 Hubble types  as follows: 
S0a and earlier ($T>$0); Sa, Sab (0$<T<$2); Sb, 
Sbc (2$<T<$4); Sc, Scd (4$<T<$6); Sd and later ($T>$6).
The vertical arrows mark subsample medians, calculated with a 
type index resolution of unity. 
(Figure is from Hunt \& Malkan 1999)
}
\label{fig:1}       % Give a unique label
\end{figure}

Do local Seyferts and central starbursts  
reside preferentially in certain type of galaxies? 
Using  the  12 $\mu$m sample (E12GS) 
and the CfA sample of Seyferts,
Hunt \& Malkan (1999; Fig. 3) report that 
\it 
Sy 1 and Sy 2 
nuclei tend to reside primarily in early-type (E--Sbc)
galaxies.  
\rm
The Hubble type quoted here is the RC3 Hubble index based 
on visual classifications of optical images.
A similar result on Seyferts  is reported by Ho et al. (1997a) 
from the POSS optical spectroscopic survey 
which tends to include lower luminosity galaxies 
and has a median extinction-corrected H$\alpha$  
luminosity of  only $2 \times 10^{39}$ erg s$^{-1}$.
These findings on Seyferts are consistent with
earlier less comprehensive studies (e.g., Hummel et al. 1990; 
Terlevich, Melnick, \& Moles 1987;
Balick \& Heckman 1982).
H{\sc ii} host galaxies tend to have  later Hubble types 
than Seyferts according to both E12GS (Hunt \& Malkan 
1999) and POSS (Ho et al. 1997a), but the mean value of 
the Hubble type varies in the surveys, possibly due to
luminosity differences.

Properties of AGN hosts in the redshift range 0.5--2.5
are particularly interesting as the optical QSO
activity peaks at $z\sim$ 2.5.  Keen insights 
are stemming  from two  large panchromatic $HST$  
surveys, the  Galaxy Evolution from Morphology and SEDs 
(GEMS; Rix et al. 2004) 
and the Great Observatories Origins Deep Surveys 
(GOODS; Giavalisco et al. 2004).
A GEMS study of 15 AGN which have $M_{\rm B}$ $\simeq$ -23 
and are in the redshift range  0.5$<z<$1.1 
where comparable data for control inactive galaxies 
exist, report that 80\% of the AGN hosts are early-type (bulge-dominated)  
compared to only 20\% that are disk-dominated (Sanchez et al. 
2004). 
The high  rest-frame $B$-band concentration indices of the  GOODS 
AGN  at $z\sim$~0.4--1.3 (Grogin et al. 2004; $\S$ 2.2)  also support 
the interpretation that these systems are predominantly 
bulge-dominated.
% The morphological typing   is based  on single Sersic 
% fits as well as exponential  and  de Vaucouleurs fits.

Furthermore, Sanchez et al. (2004) report that a much larger  
fraction  (70\%) of the early-type AGN  hosts at   0.5$<z<$1.1 
show  blue global rest-frame  
$U$-$V$ colors, compared to inactive early-type galaxies 
in this redshift and luminosity range.
These global  blue colors are   consistent with the presence of young 
%(0.3--3 Gyr with an mean age of 1 Gyr) 
stellar  populations over large regions of the AGN host galaxies.
The trend of enhanced blue colors in AGN hosts at  0.5$<z<$1.1 
seems to hold  both at higher and lower redshifts.
SDSS spectra of local  $z<0.2$  Sy2 galaxies show 
a significant contribution from young stellar populations, 
and this trend is strongly correlated with nuclear 
luminosity (Kauffman et al. 2003).
At  higher redshifts (1.8 $<z<$ 2.75), Jahnke et al. 
(2004)  find that the host galaxies of 9  moderately bright 
($M_{\rm B}$ $\simeq$ -23) AGN   in the GEMS survey have 
rest-frame UV colors that are considerably bluer than expected 
from an old population of stars. Unfortunately, for these 9  
distant AGN the detection images are not deep enough 
to constrain the morphology. 
Earlier studies of a handful of luminous QSO at $z$~$>$~2 also 
reported very UV-luminous hosts (e.g., Lehnert et al. 1992; 
Hutchings et al. 2002).

One possible interpretation of the enhanced global blue  colors
exhibited by AGN hosts is that the mechanism which ignites 
the central BH in these galaxies also triggers global SF. 
The fact that SF is  triggered not only in the nuclear
region, but over an extended (several kpc) region, would 
tend to exclude major (1:1) mergers  and favor weaker  (e.g., 
minor (1:10) or intermediate mass-ratio (3:1))  mergers/interactions 
where the gas has a larger $L$, and typically 
settles in extended inner disks during simulations (see $\S$ 5). 
In fact, only  3/15 of the  AGN hosts in the Sanchez 
et al. (2004) study show  signs  of strong disturbances.  
%and at most 6/15 show any visible signs of tidal interactions.
In the same vein, the GOODS AGN study (Grogin et al. 2004; $\S$ 2.2) 
reports no significant difference between the  rest-frame $B$ 
asymmetry index of active and inactive galaxies  over $z\sim$~0.4--1.3.
This  suggests that AGN do not preferentially occur in major mergers 
over this redshift range.

\section{Interactions and AGN/Starburst Activity }

\subsection{Basic Physics of Major Mergers}

The term 'major merger' usually refers to the merger of two 
disk galaxies with a mass ratio  of order 1:1.
Simulations of major mergers (e.g., 
Negroponte \& White 1983; 
Noguchi 1988; 
Barnes \& Hernquist 1991; 
Heller \& Shlosman 1994; 
Mihos \& Hernquist 96; Struck 1997) 
show that they generate large gas inflows into the inner kpc
and could plausibly trigger intense starbursts and AGN 
activity. The full parameter space controlling the outcome of major
mergers has not yet been fully explored, but  I summarize here 
(see also Fig. 2)  some  salient general findings:

\begin{enumerate}
\item
Not all  speeds, energies, angular momenta, and orientations are 
equally effective in inducing large gas inflows, rapid 
mergers, and  disruptions  during a major merger.
For instance, while all bound orbits will eventually lead to mergers, 
low  angular momentum and low energy orbits will lead to more rapid 
mergers. 
Prograde mergers, where  the spin and orbital angular momenta are
aligned, occur faster than retrograde mergers, lead to more violent 
disruption, and excite larger non-circular motions (e.g., Binney 
\& Tremaine 1987). 
\item
%Orbital angular momentum  gets converted to spin angular momentum
%via tidal interactions and shocks.
Hydrodynamical  torques (shocks) tend to be important 
in the initial collision when they add spin angular momentum to the 
gas in both disks (Mihos \& Hernquist 1996; Barnes \& Hernquist 1996), 
but gravitational torques dominate thereafter.
% They also tend to be smaller than subsequent gravitational torques.
\item
In the  \it early stages \rm of the merger, 
gas inflows  primarily result 
from
\it
gravitational torques exerted by a stellar bar
\rm
induced in the disk of the
two galaxies (e.g., Mihos \& Hernquist 1996; Heller \& Shlosman 1994; 
Noguchi 1988, Sellwood 1988; Hernquist 1989). 
The torque from the bar 
is significantly larger  than the gravitational torque
exerted by the galaxies on each other. Thus,  
early gas inflows are primarily the result of gas response
in a barred potential as outlined in $\S$ 6.
Mihos \& Hernquist (1996) find that the strength of the bar 
induced decreases as the   bulge-to-disk ($B/D$)
ratio of the disk increases. This behavior is consistent 
with the idea that a  dynamically hot bulge  
component  stabilizes a disk against a bar mode.  
\item
In the  \it final  stages \rm of the merger,  as the 
galaxies merge and undergo violent relaxation, 
the gas experiences  rapidly varying   gravitational torques  
as well as shocks on interacting  orbits  (e.g., Mihos \& Hernquist 
1996), and therefore loses energy and angular momentum.
Large gas inflows ($\gg$ 1M$_{\odot}$ yr$^{-1}$)
can result in the late merger stages, provided
the episodes of SF triggered by earlier inflows have not 
depleted most of the gas already. 
 
In simulations with highly  gas-rich disks (Heller \& Shlosman 1994), 
large gas concentrations  build up in the inner kpc leading 
to the formation of self-gravitating gas clumps which are 
driven to smaller scales by  dynamical friction from the 
stellar background  (see also $\S$ 9).  Gaseous nuclear bars 
form via gravitational instabilities and drive 
further inflow.

% \item
% Mihos \& Hernquist (1996) suggest that the bulge-to-disk ($B/D$)
% ratio of the merging disk galaxies plays a determining role in  
% the timing and intensity of  the largest gas inflow and central
% starburst.  
%  For galaxies with dense central bulges, the bar induced
% in the early stages is weak because a dynamically hot bulge  
% component  stabilizes a disk against a bar mode. 
% Thus, in these systems the early stages of the merger 
% are characterized by modest inflow rates and 
% central SF, thereby leaving available large reservoirs of gas  
% to power an intense burst in the final merger phases, as is often
% seen in many ultra luminous infrared galaxies (ULIRGs).
% Conversely bulgeless or low $B/D$  disks have strong induced
% bars, large  inflow rates and powerful starbursts in the early
% phases, leading to only modest SF in the final merging phase.

\item 
After violent relaxation, the end-product of  a major merger 
tends to have an  r$^{1/4}$ de Vaucouleurs-type  stellar profile and 
boxy isophotes  similar to many luminous elliptical galaxies. 
In the case of intermediate mass ratio (e.g., 1:3)   mergers, 
the gas has  a larger specific angular momentum and tends to 
settle into an extended inner disk (Naab \& Burkert 2001)  
rather than being as centrally concentrated as in a 1:1 merger. 
The stellar component  has an  r$^{1/4}$ profile, 
disky isophotes, and isotropic  velocity dispersion similar to
lower luminosity disky ellipticals (Naab \& Burkert 2001).

\item
Future work  has yet to fully explore the parameter space 
which can influence a merger such as 
the  energy and orbital angular momentum of the initial orbits, 
% (e.g., elliptic, hyperbolic orbits),
the relative  alignment of the spin and orbital angular momentum, 
the orbital geometry, the  gas content, and the mass ratios.  
Another key step is to realistically incorporate in simulations 
feedback effects from SF and thermal cooling effects which
lead to a multi-phase ISM (e.g., Struck 1997; Wada \& Norman 2001).
Shocks from SNe can dissipate orbital energy  and transfer 
angular momentum outwards. 
Major heating  processes associated with SF 
such as stellar winds, SNe, and UV photoheating  are sources of 
angular momentum, leading to fountain flows 
and starburst-driven winds (Jogee, Kenney, \& Smith 1998; 
Heckman et al. 1990).
% Hydrodynamical simulations which include gas dynamical effects (rotation, 
% shear, shocks), self-gravity, thermal cooling   effects, and
%  feedback from SF   (Wada & Norman 1999, 2001; Wada 2001) 
%  produce a multi-phase ISM which naturally has  filaments, fine 
%  structures, clumps and holes.

\end{enumerate}

\subsection{Basic Physics of Minor Mergers}

The term 'minor merger' usually refers to 
the merger between a large disk galaxy
and  a satellite with  a mass ratio of order 1:10.  Such 
mergers  are believed to be very common (Ostriker \&  Tremaine 
1975) and  have been the subject of numerous simulations (e.g.,
 Hernquist \& Mihos 1995; Mihos et al. 1995; Quinn, Hernquist,
\& Fullagar 1993; Walker, Mihos, \&  Hernquist 1996). 
A few of the general principles  underlying this class of 
mergers are outlined below. 
	
\begin{enumerate}
\item
As it moves through the  dark matter halo, 
the satellite  experiences dynamical friction  
and sinks rapidly towards the main disk on a timescale 
$t_{\rm df}$ $\propto$ ($R^2$ $v/M$ ln$\Lambda$),  
where $M$ is the mass of the satellite, $v$ is its 
speed,  $R$ is the galactocentric radius, and 
ln$\Lambda$ is the Coulomb logarithm (Binney \& Tremaine 1987). 
For typical values of these parameters, 
the dynamical friction timescale is comparable 
to a few orbital periods or a few Gyr. The orbital angular momentum 
of the satellite is converted into spin angular momentum 
for the disk and halo of the main galaxy. 

\item
The satellite first tends to sink in the disk of the primary
where it drives warps before it sinks towards the central 
regions (Quinn et al. 93; Hernquist \& Mihos 1995). 
Depending on the shape of the halo, these warps 
can persist for a significant fraction of time before they 
are washed out by phase mixing (Quinn et al. 93; Dubinski 1994). 

\item
The satellite exerts tidal torques on the main  stellar disk 
and 
\it  induces in it large amplitude non-axisymmetries  
\rm
(Hernquist \& Mihos 1995; Mihos et al. 1995; Quinn et al. 1993) 
such as stellar spirals and  bars.  Since gas  is 
collisional and dissipative, 
the gas response leads the stellar response (see $\S$ 6) and
gas in the disk  is gravitationally torqued towards the 
central few hundred pc.
In effect, \it 
the gas in the main disk experiences a much larger
gravitational torque  from  stars in the disk than 
from the satellite.
\rm 
In simulations by  Hernquist \& Mihos (1995),  
a large fraction of the gas in the main
disk can be driven into the central regions in this way.

\item
The satellite may have a significant fraction of  
its material tidally stripped before it sinks towards
the inner part of the main disk.
%The tidally stripped material forms transient leading and 
%trailing streamers which wind up to form transient spiral that 
%fades in a few orbital periods.

\item 
Overall, the minor merger leads to large gas concentrations
in the inner few 100 pc which may be relevant for fueling 
starbursts and AGN.  It also heats the inner parts of disks 
vertically and increases the  disk thickness and velocity 
dispersion  of stars. Some authors  (e.g., Quinn et al. 1993; 
Walker  et al. 1996) have even suggested that minor mergers 
may be the origin  of thick disks.

\end{enumerate}

\subsection{Correlations between Interactions and Starbursts}

Colorful examples abound of  starbursts  and AGN  activity 
occurring in  interacting or merging galaxies. 
However, statistically significant correlations between 
central activity and  signs of morphological disturbance 
have only been found in the case of highly luminous 
starbursts or AGN. 
Signs of strong tidal interactions and 
mergers are ubiquitous in  systems where the  SR rate (SFR) 
is estimated  to be $\ge$ 10 M$_{\odot}$ yr$^{-1}$ 
such as ultra-luminous infrared galaxies (ULIRGs;  Veilleux, 
Kim, \& Sanders 2001; Sanders \& Mirabel 1996), 
and the brightest  Arp galaxies (Hummel et al. 1990; 
Kennicutt et al. 1987). 
In the local Universe, more than  95\% of ULIRGS in the 1 Jy 
IRAS sample show optical and NIR morphological signatures  of a strong 
interaction or merger in the form of tidal tails, bridges, 
double nuclei, and overlapping disks (Veilleux et al. 2001).
In a study based on  panchromatic $HST$  GOODS data,  
optically-selected bright starburst galaxies at $z \sim$~1 
show a larger frequency ($\sim$ 50\%) of disturbed 
and asymmetric  morphologies, compared to 13\% and 27\%, respectively,  
in control samples of early-type and late-type 
galaxies (Jogee et al. 2003; Mobasher et al. 2004).

While the most extreme \it individual \rm starbursts  (in 
terms of luminosity or luminosity per unit mass of gas) 
may be triggered by a major or intermediate mass ratio 
interaction, it is important to bear in mind that  
the \it cumulative \rm SFR density at a given cosmic 
epoch may be  dominated by SF emanating from a large number of 
relatively undisturbed galaxies rather than from  SF in major mergers.  
This is a particularly viable possibility out to intermediate 
redshifts  ($z \sim$~1) where major mergers are relatively  rare. 
In fact, preliminary findings  (Wolf et al. in prep.) 
from a study of $\sim$~1400 galaxies at $z\sim$~0.7 
in the GEMS survey suggest 
that  while  galaxies with strongly disturbed  rest-frame
optical morphologies are amongst the most  UV-luminous candidates, 
they only make up  a small fraction of the  UV luminosity 
density  (uncorrected for extinction) at $z\sim$~0.7. 
While this result on UV luminosity density does not 
directly translate to SFR density due to potential 
extinction effects, it  does highlight the need for 
further studies on how mergers/interactions impact 
central activity.

\subsection{Correlations between Interactions and AGN}

An excess of companions  
in local Seyferts has been reported in early papers (e.g., 
Dahari  1984; Keel et al. 1985), but more recent studies with large
or/and well-matched control samples show no strong correlations 
(e.g., Schmitt 2001; Laurikainen \& Salo 1995). 
Furthermore, in a study of 69 galaxies belonging to 31  Hickson 
compact groups (CGs), Shimada et al. (2000) find no difference between  
the frequency of H{\sc ii}  or AGN  nuclei in  Hickson CGs and field 
galaxies.

At intermediate redshifts ($z\sim$~0.4--1.3), studies based on 
GOODS data  (Grogin et al. 2004; see $\S$ 2.2) report  
no significant difference between the  rest-frame $B$ 
asymmetry index of AGN hosts  and inactive galaxies. 
This suggests  that over this redshift range, AGN hosts 
are not  preferentially  major mergers. Similarly, in the GEMS 
sample of moderately bright  ($M_{\rm B}$ $\simeq$ -23)   AGN  
at  0.5$<z<$1.1 studied by Sanchez et al. (2004; see $\S$ 4), 
only 3/15 of the  AGN hosts signs  of strong disturbances,  
and at most three others are associated with nearby potentially 
interacting companions.

\vspace{0.1 cm}
In fact,  statistically significant correlations between AGN 
activity and  signs of strong interactions 
% or nearby  companions  
are  reported only in systems with high mass accretion 
rates ($\ge$ 10  M$_{\odot}$ yr$^{-1}$) such as  very 
luminous or radio-loud QSOs (Disney at al. 1995; Bahcall et al. 
1997; Kirhakos et al. 1999),  
and  FR-II radio galaxies (Hutchings 1987, Yates et al. 1989).
I discuss below  possible  reasons as to  why a  correlation 
is only seen at the very high luminosity end: 
\begin{description}
\vspace{-0.1cm}
\item
$\bullet$
It is possible that the large mass accretion rates 
($\ge$ 10  M$_{\odot}$ yr$^{-1}$)   on small scales   required
by  very high luminosity  QSOs are realized  in nature primarily 
during violent processes such as some classes of  
major/intermediate mass-ratio interactions. 
In particular, the  last phases of a major merger 
produce  rapidly varying  gravitational torques and  strong  
shocks on intersecting orbits (see $\S$ 5.1) that are expected
to generate large  inflow rates on small scales.

\item
$\bullet$
Conversely,  in order  to fuel  Seyferts over their nominal  
duty cycles (10$^{8}$ years)  with a  mass inflow rates $\le$ 0.01  
M$_{\odot}$ yr$^{-1}$), any one of the many  10$^{6}$ M$_{\odot}$  
clouds commonly present within the inner 200 pc radius is adequate, 
provided  that some fueling mechanism  drains 
its  angular momentum by more than 99.99\%  (Fig. 2). 
Given that the required fueling  mass  is
\it 
10$^{-3}$--10$^{-2}$ 
\rm 
of the amount of gas typically  found in the inner few 100 pc, 
it is well  possible that  \it localized  \rm   low-energy processes  
(e.g.,  SNe shocks on ambient  clouds)  may be adequate for reducing 
its angular momentum. In other words, interactions and other fueling  
mechanisms which affect the  \it bulk \rm of the gas may not be 
necessary  for fueling Seyferts.

\end{description}

\section{ Large-Scale Bars  in Starbursts/AGN  Hosts}

\subsection{Bar-Driven Gas Inflow: Theory and Observations}

\begin{figure}
\centering
%\includegraphics[height=4.5cm]{Rings78a.eps}
%\includegraphics[height=4.5cm]{Rings79a.eps}
\includegraphics[height=4.5cm]{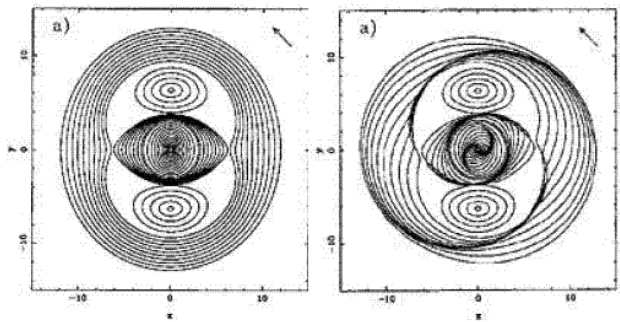}
\caption{ 
\bf Stars and gas in a barred potential:  \rm 
\it 
Left:  
\rm
The periodic \it stellar \rm orbits  in the potential 
% % or  cos2$\theta$
of a bar  (oriented horizontally here) 
change orientation by $\pi$/2 at each resonance.
The $x_1$ family of periodic stellar orbits is elongated along
the bar major axis and supports the bar.
The $x_2$ orbits are elongated along the bar minor axis and 
and exist inside the ILR or between the ILRs. 
\it
Right:
\rm 
In contrast, gas being collisional and dissipative, follows orbits 
which  change their orientation  only gradually,
leading to spiral-shaped gas streamlines.
(Figures are adapted from Buta \& Combes 1996)
}
\label{fig:1}       % Give a unique label
\end{figure}

Bars are ubiquitous in most  ($>$ 70\%)  local spiral galaxies 
(Grosb{\o}l et al. 2002; Eskridge et al. 2002), 
and recent work (Jogee et al. 2004a,b; Sheth et  al. 03)  suggest 
they may be quite abundant  out to $z \sim$~1. 
Bars drive the dynamical
evolution of disk galaxies by exerting gravitational torques which 
redistribute mass and angular momentum. 
In fact, large gas inflows into  the inner few hundred pc of a disk galaxy  
primarily result from gravitational torques exerted 
by a  stellar bar. This is true not only in the case of an 
isolated barred  galaxy, but also for  some classes  of 
minor mergers ($\S$ 5.2; Quinn et al. 1993; 
Hernquist \& Mihos 1995; Mihos et al. 
1995), most intermediate 1:3 mass ratio mergers 
(Naab \&  Burkert  2001), and  the early phases of most  major mergers  
($\S$ 5.1; Noguchi 1988, Sellwood 1988; Hernquist 1989; 
Heller \& Shlosman 1994; Mihos \& Hernquist 1996). 
I will, therefore, devote a fair share  of this review to a 
discussion of the  basic principles of  bar-driven gas inflow, 
what it can achieve in the context of fueling starbursts and AGN, 
and related observations.

A barred potential is made up of different families of periodic stellar
orbits which conserve the Jacobi integral ($E_{\rm J}$), a combination
of energy and angular momentum (e.g., Binney \& Tremaine 1987). 
The most important families are those oriented parallel to  the 
bar major axis ($x_1$ orbits; Contopoulos \& Papayannopoulos 1980) 
and  minor axis ($x_2$ orbits). The $x_1$ 
family is the main family supporting the bar and 
extends between the center and the corotation resonance (CR) of the bar. 
The inner Lindblad resonances  demarcate the transition region 
where the dominant family of periodic stellar orbits changes from 
$x_1$  to $x_2$.  Thus, the $x_2$ family appears  between the center
and the ILR if a single ILR  exists, and between the inner ILR 
(IILR) and the outer ILR (OILR) if two ILRs exist. 
%The orientation of populated periodic orbits changes by $\pi/2$ 
%at each  resonance  (Fig 2a).                        
% When the bar is strong enough, the x2 orbits disappear. 
% The bar strength necessary to eliminate the x2 family depends on the 
% pattern speed Omega b: the lower this speed, the stronger the bar must be.
Gas tries to follow these orbits, but  due to its collisional and 
dissipative nature, it cannot remain on periodic orbits which cross.
Instead, the gas-populated orbits change their orientation 
only gradually due to shocks induced by the finite gas pressure, 
leading to spiral-shaped gas streamlines, offset with respect
to the stars  (Fig. 4).
% periodic orbits are crowded and may self intersect,
% so gas shocks on leading side of stellar bar e.g., Athanassoula 92
% The spiral gas streamline tends to turn by nearly 90 $\deg$
% each resonance and  can change shape from trailing to leading  
% towards the central region of the galaxy as the radial gradient  
% of $\Omega$ - $\kappa$/2  changes from negative to positive. 
%Since the spiral gas streamlines  are offset with respect
%to the stars, they  experience a  net  gravitational 
%torque  from the stars.  
% The direction of the torque depends on the orientation of the gas
% streamlines with respect to the bar major axis and the location
% within the bar potential.
The negative torque exterted by the stars 
between CR and the OILR causes the
gas to lose angular momentum and flow towards the inner kpc.

\begin{figure}
\centering
\vspace{1cm}
\includegraphics[width=8cm]{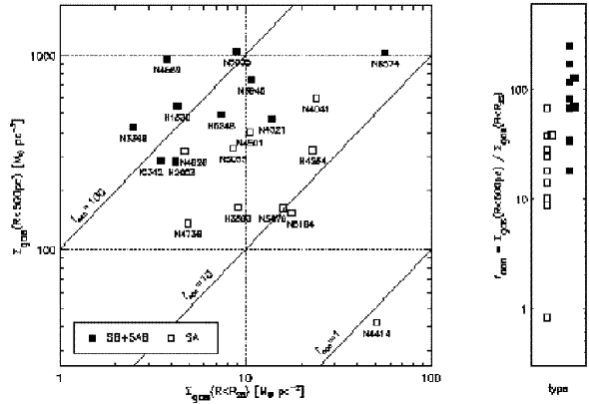}
\caption{ 
\bf Molecular gas central concentrations in barred galaxies: \rm 
\it 
Left:
\rm 
Comparison of the molecular gas surface densities averaged 
within the central 500 pc radius and over the optical 
de Vaucouleurs radius ($R_{\rm 25}$).
The spatial resolution of the data is $\sim~4\arcsec$. 
The ratio   of these two quantities  is a 
measure of the molecular gas  central concentration ($f_{\rm con}$) 
within the central kpc.  
\it 
Right: 
\rm 
On average the molecular gas  central  concentration
($f_{\rm con}$) within the central kpc is higher in barred 
galaxies (filled squares) than in unbarred galaxies 
(open squares). (Figure is adapted from Sakamoto et al. 1999)
}
\label{fig:1}       % Give a unique label
\end{figure}

\begin{figure}
\centering
\vspace{1cm}
\hspace{-2cm}
\includegraphics[height=15cm]{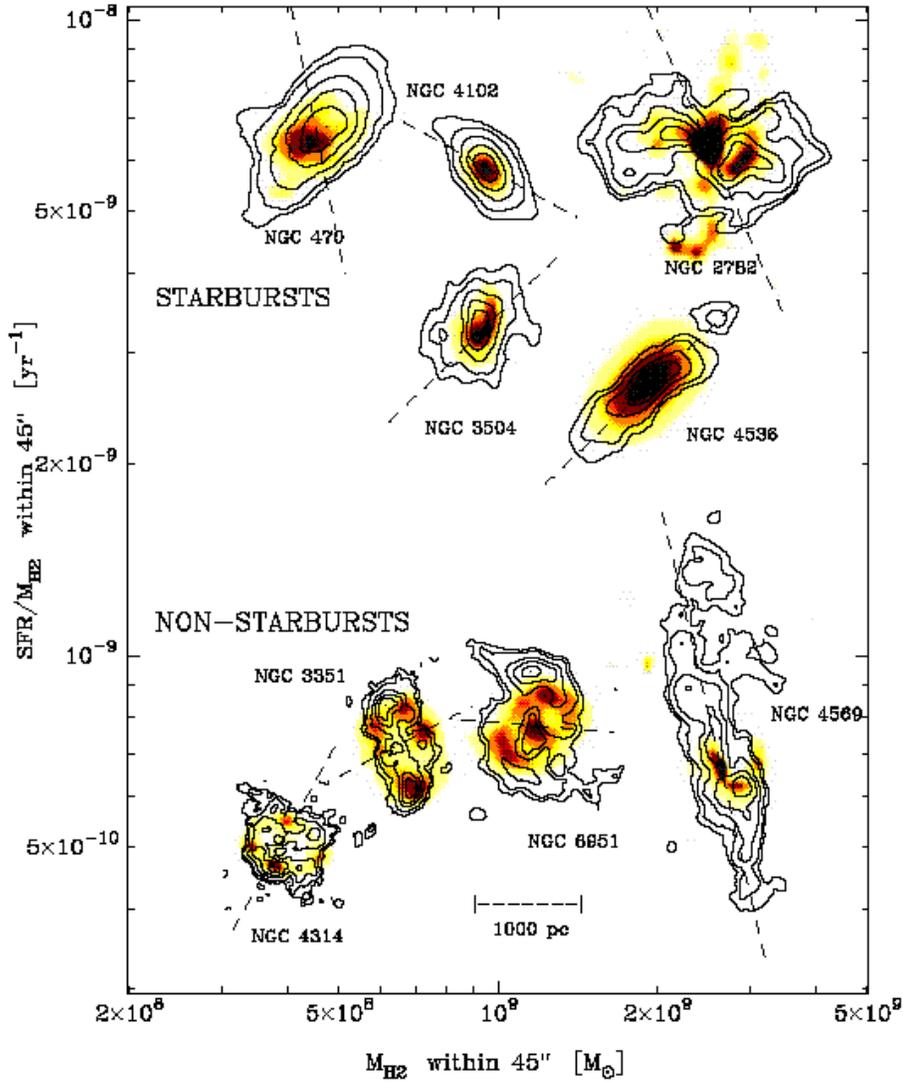}
\hspace{0cm}
\caption{ 
\bf 
High resolution ($2\arcsec$ or 100--200 pc)  
observations of molecular gas and star
formation in the inner few kpc of barred galaxies: 
\rm
CO (1--0) intensity  (contours)  maps  of the circumnuclear 
region of  barred galaxies are  overlaid on 
1.5 or 4.9 GHz radio continuum map  or H$\alpha$  map (greyscale). 
The dotted line is  the P.A. of the  large-scale stellar bar/oval.
A wide range in gas distributions and in SFR per unit mass  
of molecular gas (SFR/$M_{\tiny \rm H2}$) is present.  
Systems with high/low circumnuclear SFR/$M_{\tiny \rm H2}$
are denoted as starbursts/non-starbursts.  
Part of the range in SFR/$M_{\tiny \rm H2}$ and the variety of 
gas distributions can be understood in terms of different stages 
of bar-driven inflow and the critical density for the onset of SF. 
(From Jogee et  al. 2004c; see text for details).
%In some candidates, the nuclear
%regions  also show LINER (NGC 4314) and Seyfert 2 (NGC 6951) 
%activity. 
}
\label{fig:1}       % Give a unique label
\end{figure}

There is mounting observational evidence for bar-driven gas inflow 
into the inner kpc.
Moderate ($4\arcsec$) resolution  CO($J$=1--0) interferometric 
surveys of molecular
gas in the circumnuclear  region of nearby spirals show that 
the molecular gas central  concentration   (Fig. 5)
is on average higher  in barred than in unbarred galaxies  
(Sakamoto et al. 1999). 
Barred galaxies also show shallower metallicity gradients across 
their galactic disks  than  unbarred ones (Martin \& Roy 1994; 
Vila-Costas \& Edmunds 1992). 
Observations of cold or ionized gas velocity fields show   evidence for shocks and 
non-circular  motions along  the large-scale 
stellar bar  
(Quillen et al. 1995; Benedict, Smith , \&  Kenney 1996; 
Regan, Vogel, \& Teuben 1997; Jogee  1999;  Jogee, Scoville, \& Kenney 2004c).
Bar-driven gas inflow rates  into the inner kpc 
have been estimated only in the case of a  few strong bars and  
range from 1 to 4 M$_{\tiny \odot}$ yr$^{-1}$ 
(Quillen et al. 1995; Regan et al. 1997; 
Laine, Shlosman, \& Heller 1998).
% based on various methods including estimates of the gravitational 
%torque at different points along the bar  using 
%near-IR and  CO observations (Quillen et al. 1995), and  comparisons 
%of observations of  with SPH and hydrodynamical 
% simulations 

\begin{figure}
\centering
%\vspace{1cm}
%\includegraphics [width=7cm]{plotsam2.FIG2.den.wo3310.sam2.eps}
\includegraphics [width=7cm]{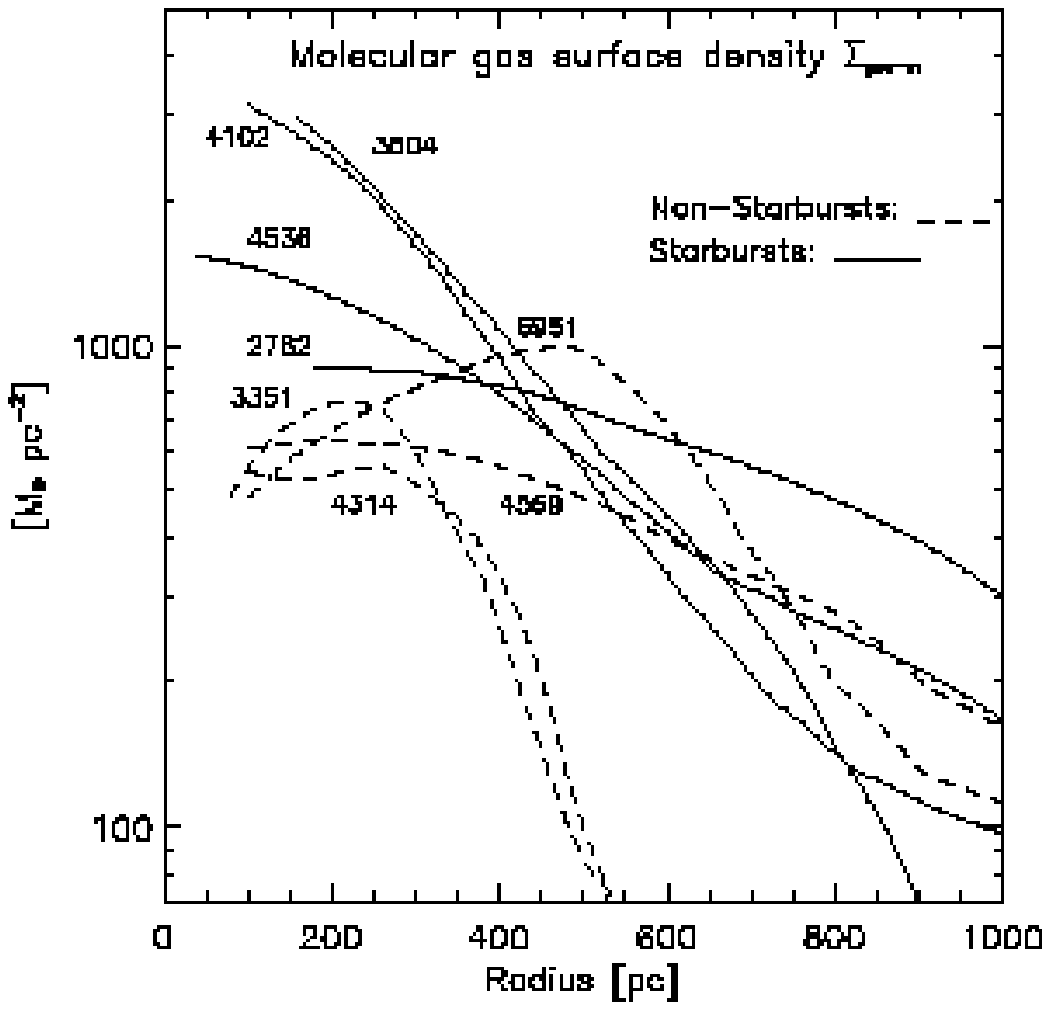}
\caption{ 
\bf
Molecular gas surface densities in the inner kpc of barred galaxies: 
\rm
The azimuthally averaged molecular gas surface density derived 
from the high resolution ($2\arcsec$ or 100--200 pc) CO ($J$=1--0) 
maps  in Fig. 6 are plotted.
Quantities are plotted starting at a radius $\ge$ half 
the size of the synthesized beam.
%so that  meaningful values are displayed. 
Most of the  starbursts have developed larger 
molecular gas surface densities (1000-3500 $M_{\tiny \sun}$ 
pc$^{-2}$) in the inner 500 pc radius  than the  
non-starbursts   for a given CO-to-H$_{\rm 2}$ conversion factor. 
(From Jogee et al. 2004c)
}
\label{fig:1}       % Give a unique label
\end{figure}

What happens once gas reaches the central kpc region of barred galaxies? 
High resolution (1--2$\arcsec$; 100-200 pc) observations of 
molecular gas and SF in the central few  kpc 
of eleven barred galaxies (Jogee et al. 2004c; Fig. 6) 
reveal a variety of gas distributions  and a large range in 
SFR per unit mass  of molecular gas (SFR/$M_{\tiny \rm H2}$).
These differences  in gas distributions and SF 
efficiencies  can be partly understood in terms of different 
stages  of bar-driven inflow and the existence of a critical 
density for the onset of SF (Jogee et al. 2004c). 
Some systems  appear in the early stages of bar-driven 
inflow  where  a large fraction of the gas  is still extended along 
the large-scale bar, shows large non-circular kinematics and does 
not form stars efficiently.  Other galaxies have  developed large 
gas surface densities  (600-3500 M$_{\odot}$ pc$^{-2}$; Fig 7) 
inside the OILR  of the  bar. 
Intense  starbursts appear to be triggered only once gas densities 
approach or exceed the Toomre  critical density for the onset
of gravitational instabilities (Elmegreen 1994; Jogee et al. 2004c).

While the large-scale bar efficiently drives gas from the outer disk 
into the inner kpc, theory suggests that the radial inflow of gas slows 
as it crosses the  ILR  because shocks  associated 
with the  large-scale  bar weaken,  the gravitational potential becomes
more axisymmetric, and  gravitational torques on 
the gas in the vicinity of  ILRs weaken or even reverse  
(e.g., Schwarz 1984; Combes \& Gerin 1985; Shlosman et al. 1989; 
Athanassoula 1992).  
In support of this picture, gas rings near the ILRs have also been 
reported  in many individual galaxies    (e.g., 
Kenney et al. 1992;  Knapen et al. 1995; Jogee 1999; Jogee  et al. 2001).
Figure 8 illustrates the results from  high resolution  
observations of gas in the centers of barred galaxies  
(Jogee et al. 2004c). 
In the seven barred galaxies shown,  typically the OILR radius is 
$\ge$  500 pc while the IILR radius is $\le$ 200--300 pc.
It should be noted that even with these high resolution (100 pc) 
data, it remains difficult to assess the presence/response
of gas  \it inside \rm the IILR since the latter  radius is 
comparable to the spatial resolution.

\begin{figure}
\centering
\hspace{-1.71cm}
\includegraphics[height=6.1cm]{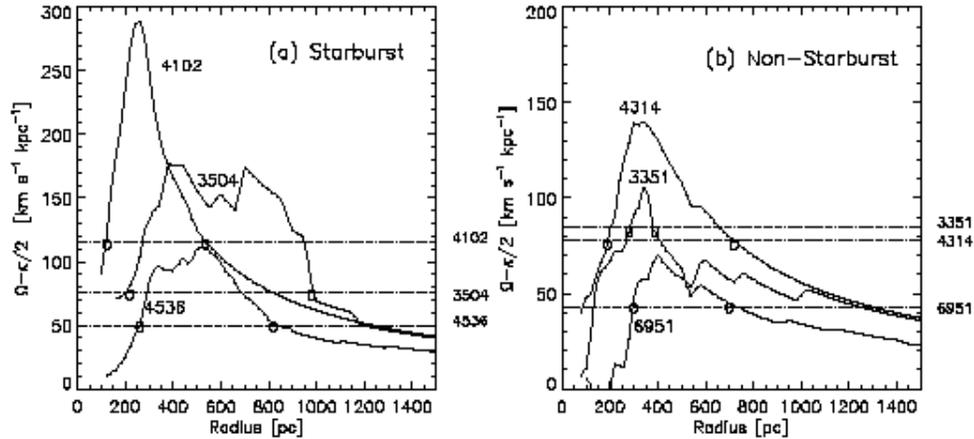}
\caption{ 
\bf
Location of Inner Lindblad resonances in barred galaxies: 
\rm
[$\Omega$ - $\kappa$/2] is plotted against radius 
for the circumnuclear region of the barred  galaxies 
shown in Fig 6. 
Under the  epicyclic approximation valid for weak bars, the intersection 
of [$\Omega$ - $\kappa$/2] with $\Omega_{\rm p}$ defines the 
locations of the ILRs of the large-scale stellar bar.
The upper limit on the  bar pattern speed ($\Omega_{\rm p}$)  
is drawn as a horizontal  line and  is estimated  by assuming 
that the corotation resonance is at or beyond  the bar end.  
Values range from 43 to 110 km~s$^{-1}$  kpc$^{-1}$.
%The rotation curves are
%derived from high resolution (100-200 pc) CO (1-0) interferometric 
%observations. 
Typically, the OILR radius is $\ge$  500 pc while the IILR radius 
is $\le$ 200--300 pc. 
Thus, in these systems, the large molecular gas densities (Fig. 5) 
have  developed inside the OILR  of the bar/oval.
% It is unclear if the gas is inside the inner ILR  (IILR) 
% as the latter's estimated location is comparable to 
% the resolution of the observations. 
(From Jogee et al. 2004c)
}
\label{fig:1}       % Give a unique label
\end{figure}

The question of  whether bars are long-lived or whether they 
dissolve and  reform recurrently  over a Hubble time
is  highly controversial  at the  present time. 
Early studies  (Hasan \& Norman 1990; 
Friedli \& Benz  1993; Norman, Sellwood, \& Hasan 1996, but
see also  Bournaud \& Combes 2002; Shen \& Sellwood 2004) 
proposed that  once a  large  
central mass concentration (CMC) builds up in the inner 100 pc 
of a galaxy, it will destroy or weaken the bar 
due to the development of chaotic orbits and reduction of
bar-supporting orbits. 
In some scenarios, bars  even cause their own demise and 
self-destruction by  building up  such  CMCs via gas inflows. 
The disk left behind after a bar is destroyed 
is dynamically  hot and does not reform a new bar 
unless it is cooled significantly.  
Recently, Bournaud \& Combes (2002) suggested that  
bars are destroyed primarily due to the reciprocal torques 
of gas on the stars in the bar (rather than by the CMC \it per se \rm). 
Within this framework, bars can dissolve and reform  recurrently if the 
galaxy accretes  sufficient cold gas over a Hubble time.
In fact, Shen \& Sellwood (2004)  find  that  in purely stellar
$N$-body simulations, the bar is  quite robust to CMCs.
% Even the
% most destructive point-like CMCs need to be a few \% of the total 
% disk mass to completely destroy the bar over a Hubble time, while 
% diffuse CMCs require more than 10\% of the disk mass. 
Ongoing studies (Jogee et al. 2004a,b) of the bar properties 
and CMCs of galaxies over lookback times of 9 Gyr 
(out to $z\sim$~1), based on the GEMS survey,  will help 
provide discriminant tests on the evolution and  lifetime of bars 
out to  $z\sim$~1.

\subsection {Correlations between Large-Scale Bars and  Starbursts}

Analyses of the  extended 12 $\mu$m Galaxy sample (E12GS) 
show that the fraction of large-scale bars  is larger in 
starburst nuclei (82--85\%) than in  normal (61-68\%) ones (Fig 9; 
Hunt \& Malkan 1999).
The term  'normal' here denotes quiescent nuclei which do not have 
HII/starburst, LINER, Sy 1,  or Sy 2 signatures.
Similar correlations  between large-scale bars and circumnuclear 
starbursts  were previously reported by  studies with smaller 
or/and less complete samples.
For instance, Hawarden et al. (1986) found  that a large fraction 
of SB and SAB galaxies  show an enhanced  IRAS 25 $\mu$m flux, which
they assigned  to circumnuclear rings  of SF present in the central 
$20\arcsec$.
%To support the interpretation of enhanced 25  $\mu$m  flux as being
%due to active SF, 
%a) authors construct simple models using the 
%average observed flux of a group of luminous unresolved HII 
%regions in Cygnus.  Adding 10$^{\rm 5}$ such objects to  a 
%system with mean IRAS 
%color and lum of  SA galaxies in fig 1 reproduces the IRAS color and 
%lum of a typical 'starburst' system such as 1097. 
%Correlation holds after  excluding  Sy1 Sy2[D 22 and LINER.
Earlier work (e.g., Terlevich et al. 1987; 
Kennicutt et al. 1987) also reported correlations based 
on UV  and optical starbursts.

All the  afore-described  studies rely on  optical images
to identify bars and therefore suffer from the associated  
caveats  outlined in $\S$ 3.6. 
Taken at face value, they suggest that 
relatively luminous starbursts have a higher  frequency  of 
large-scale stellar  bars than  normal galaxies.
A natural explanation exists for the bar--starburst 
correlation.  The most powerful starbursts 
typically occur in the central few 100 pc of galaxies once 
large supercritical gas  densities (often as large as 
several 1000 M$_{\odot}$ pc$^{-2}$; Jogee et al. 2004c)  build up. 
A spontaneously or tidally induced large-scale bar
\rm  
is an ideal fueling mechanism  for luminous circumnuclear 
starbursts  because it 
efficiently drains angular momentum  from gas on exactly the right 
spatial scales (several kpc to a few hundred pc; Fig 2)  
relevant for building the pre-requisite large concentrations. 
\rm

\begin{figure}
\centering
% \vspace{1cm}
\hspace{-2 cm}
\includegraphics[width=7.5cm]{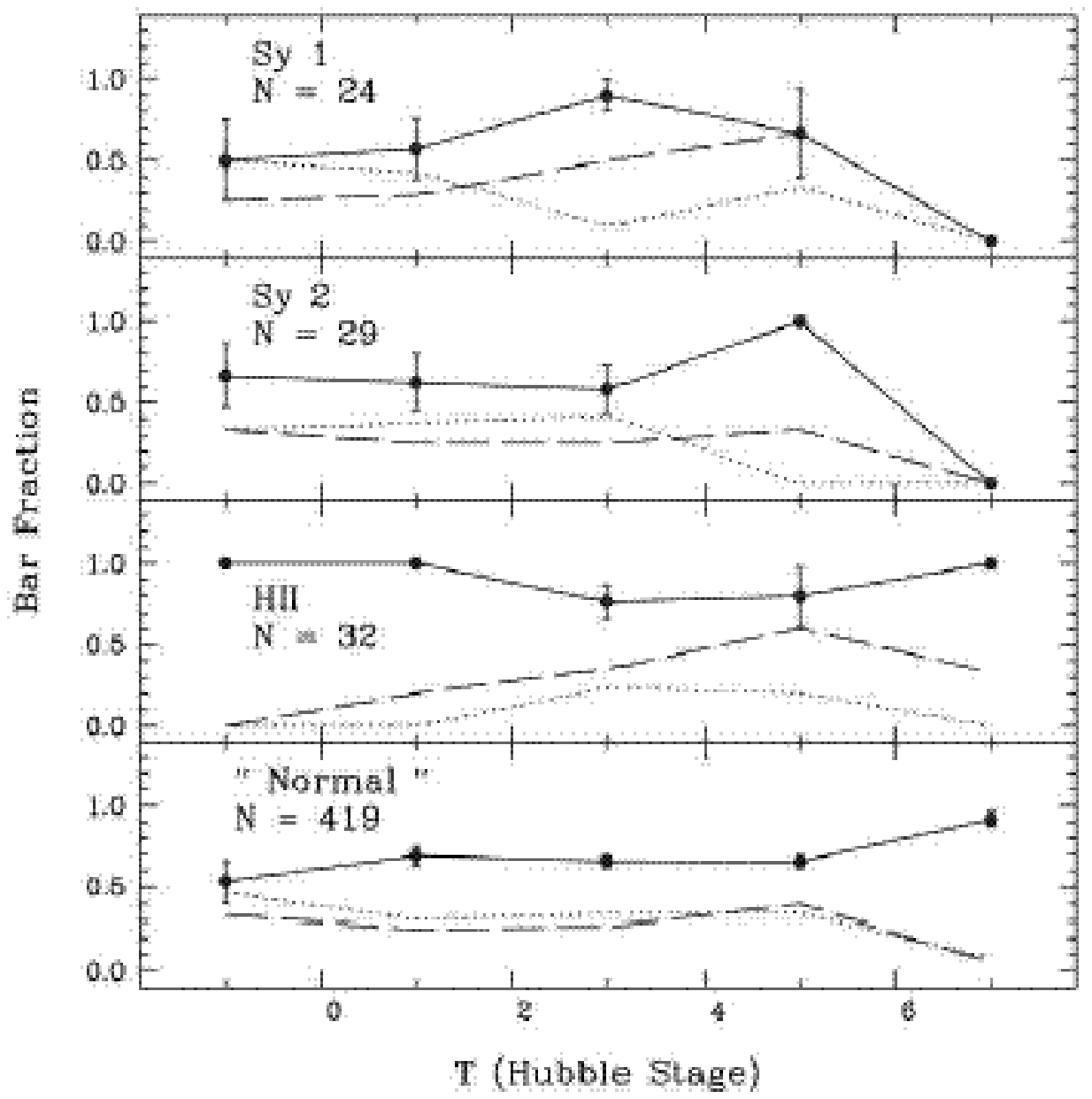}
\caption
{\bf 
Relation between large-scale bars and nuclear starburst/AGN: 
\rm
The fraction of barred galaxies  as a function of Hubble type and 
nuclear types is shown for  the extended 12 $\mu$m Galaxy Sample.  
Nuclear types (H{\sc ii}, Sy 1, Sy 2, LINER) are taken 
from NED.  Quiescent nuclei without any of these signatures are
denoted as 'normal'. Optically-based RC3 bar 
classes are shown as dotted lines (unbarred), 
dashed lines  (weakly barred SAB), and solid lines (weakly and strongly 
barred SAB+SB).  The data bins are as in Fig. 3. 
%The data have been binned in terms of RC3 Hubble types  as follows: 
% S0a and earlier ($T>$0); Sa, Sab (0$<T<$2); Sb, 
%Sbc (2$<T<$4); Sc, Scd (4$<T<$6); Sd and later ($T>$6).
% Error bars are shown only for the barred distributions, and are 
% derived from counting statistics as, where f is the fraction
%of barred objects in a given morphological type bin and N is the total 
% number of objects in the bin. 
Numbers under the panel label give the number of galaxies in each 
subsample.
% with well-defined bar classes.
(Fig. is from Hunt \& Malkan 1999)
}
\label{fig:x}       % Give a unique label
\end{figure}

\subsection {Correlations between Large-Scale Bars and AGN}
% \subsection{Large-Scale Bars  in AGN  Hosts}

The study of Hunt \& Malkan (1999) based on the E12GS sample 
and  RC3 optical bar classes finds that there is no excess 
of bars in Seyferts.  
Two recent NIR-based studies (Mulchaey \& Regan 1997; Knapen et al. 
2000) have investigated the  fraction 
$f$ of large-scale bars  in  Seyferts and   normal galaxies using 
matched control  samples, high resolution  NIR images, and ellipse 
fits to characterize bars.
Mulchaey \& Regan  (1997) report a similar incidence of bars
($f\sim$ 70\%) in  Seyferts and   normal galaxies  while 
Knapen et al. (2000) find  a  higher fraction of bars in 
Seyferts (79\% $\pm$ 8\% vs. 59\% $\pm$ 9\%) at a  significance 
level  of  2$\sigma$. 
More recently, Laurikainen, Salo, \& Buta (2004) classified bars in the
Ohio State University sample (Eskridge et al. 2002)  using  
Fourier decomposition of NIR images (Fourier bars), and  report a higher 
fraction (72\%) of such Fourier bars in Seyfert galaxies, LINERs, and 
H{\sc ii}/starburst galaxies, as compared to  55\% in the inactive 
galaxies. It is not entirely clear at this time how the 
Fourier bars identified in this study compare to bars identified 
by other methods such as ellipse fits, and  how they are impacted
by other $m$=2 modes (e.g., spirals). Furthermore, the nuclear types 
(H{\sc ii}, Seyfert, and LINERs) in the OSU sample are  not 
homogeneously classified via spectroscopic  observations.
Thus, it is fair to conclude that  at this time,  the question of 
whether Seyferts have an excess of large-scale bars compared to 
inactive galaxies remains open.

\vspace{0.1cm}
I discuss below what one might expect with regards to 
correlations between Seyferts and large-scale bars, based on theoretical  
considerations, and outline some areas where future work is needed.
% The control samples were  selected to match the Seyfert sample in 
% Hubble type,  inclination, luminosity, and other parameters.
% The Mulchaey \& Regan (1997) study finds no difference between 
% the   Seyferts and normal galaxies  while the Knapen et al (2000)
% study finds 
\vspace{-0.13cm}
\begin{enumerate}
\item
The first question one might ask is whether all  barred 
galaxies are expected to show AGN activity. 
It is clear from Fig. 2 and preceding 
discussions that a large-scale bar efficiently drives gas
only down to scales of a few 100 pc.  At that stage,  
the  specific angular momentum $L$ of the gas is still 
1000 times too high  to be digestible by a BH. Thus,
unless other mechanisms are present to reduce $L$ 
further, the gas will not fuel the central BH. 
Furthermore, even if a barred system does go through an
AGN phase,  the lifetime of a bar is expected to be at 
least  1 Gyr,  while a typical  AGN duty cycle is 10--100 
times shorter.
\it
Thus, we do not expect  all barred galaxies to show 
AGN activity.
\rm

\item
A different question is  whether  we would expect
all local Seyfert host galaxies to have a bar.
\rm 
This question is tantamount to asking whether 
we need  to transport gas 
from the outer regions (several kpc) 
to the inner few 100 pc  of a galaxy in order to  
\it directly or indirectly \rm fuel Seyferts. 
Let us first consider the required mass budget.
The  mass of gas 
required to fuel a typical Seyfert at a rate of 
10$^{-2}$ M$_{\odot}$ yr$^{-1}$ over a  nominal duty 
cycle of 10$^{8}$ years is 
\it
only  10$^{6}$ M$_{\odot}$  or 10$^{-3}$--10$^{-2}$ of 
the typical gas content 
\rm  
(10$^{8}$--10$^{9}$ M$_{\odot}$)

found in the  inner kpc of a  present-day spiral.
One might  argue, therefore, that we would not expect a strong 
correlation between Seyfert activity and strong/moderate bars 
because even weak/inefficient  large-scale  fueling  mechanisms 
are more than adequate to drive  such tiny amounts of gas 
from kpc  scales  down to  the inner few 100 pc.
Examples of such weak fueling  mechanisms would be   
oval features which are not conventionally classified as  bars,  
weak non-axisymmetries  easily induced in minor 
mergers/interactions, and  dynamical friction slowly 
sinking a gas-rich satellite  (see $\S$ 5.1).
It has in fact been argued that the strong excess of rings seen in 
Seyferts (see point 4) represent such weal oval perturbations. 
In summary, from the point of view of \it large-scale gas transport,\rm 
one would not expect  strong/moderate large-scale bars to be 
a pre-requisite for fueling Seyferts.

However, we should bear in mind that a  correlation between Seyfert 
and strong/moderate large-scale bars  might  \it indirectly \rm result  
due to  requirements for gas transport on \it small \rm scales (100s of 
pc) rather than large (kpc) scales.
Even if only  0.1\%--1\% of the gas present on scales of a few 100 pc  
provides an adequate mass budget for AGN activity over 
many duty cycles, the fueling of the BH can only occur if there
exist mechanisms on scales of a few 100 pc which can drain the angular 
momentum of this gas by more than 99.99\%. 
If some of these mechanisms, such as dynamically decoupled secondary 
nuclear bars, are favored by the presence of a moderate/strong 
large-scale bar, a  correlation between the latter and Seyferts 
might result.
However, one may also counter-argue that  with only 0.1\%--1\% 
of the circumnuclear gas present being involved in AGN fueling, 
strong nuclear fueling mechanisms such as nuclear bars may not be 
needed. Instead,  localized  \rm   low-energy processes  such as 
SNe shocks and cloud-cloud collisions may be enough to significantly 
reduce the angular momentum on one  ambient  10$^{6}$ M$_{\odot}$ 
cloud and  drive  it from 100s of pc down to tens of pc.

\item
The uncertain question of whether bars can self-destroy and
reform recurrently over a Hubble time (see $\S$ 6.1)  adds another 
pertinent dimension to the interpretation of statistical correlations 
or lack thereof between bars and Seyferts.  
%  and an increased  mass in the region of  $x_2$  stellar orbits.
% Typically,  the bar dissolves  over  a few  rotation 
% periods, typically a few  $\times 10^8$ years  
% if the  mass concentration in the central 200 pc radius 
% is $\ge$  1-2\% of the total stellar mass in the disk. 
In some models of bar destruction  (e.g., Hasan \& Norman 1990; 
Friedli \& Benz  1993; Norman, Sellwood, \& Hasan 1996), 
a bar which dutifully brings a large gas concentration into 
the inner  kpc might dissolve away   once  other 
mechanisms in the inner kpc start to relay the 
fuel to  100 pc scales  and eventually to the BH. 
Some observational support for this picture comes from 
studies  reporting that the bar strength 
%(as characterized by the ellipticity or  $Q_{\rm b}$ parameter 
% measured in NIR images) 
is weaker in Seyferts than in inactive  galaxies 
(Shlosman et al. 2000; Laurikainen et al. 2004).

\item
A  strong correlation is  seen between Seyferts  and 
the presence of large-scale rings in the host galaxies (Hunt \& Malkan 
1999): the frequency of outer rings and of  (inner +outer) rings is 
higher by 3-4  times in Seyfert galaxies 
compared to normal galaxies.
\rm 
It has been argued that this correlation may result by chance 
because both Seyferts and  rings (particularly outer rings) 
tend to be more  common in early-type systems. 
Another possibility is that these oval rings  represent  the type 
of weak non-axisymmetric distortions discussed in point 2. 
A third possibility is that a large fraction of these rings might 
be remnants of bar  dissolution. Both latter possibilities need to be 
explored further theoretically and observationally.

\item
To date, all studies between AGN and large-scale bars have focused
on local galaxies.  Yet, AGN activity  is known to  increase with 
redshift,  with the optically bright QSO phases peaking at $z \sim$~2.5 
($\S$ 3.4). It is important to extend studies of bars and AGN to 
earlier epochs. 
The ongoing work on the impact and evolution of bars over the 
last 9 Gyr  (out to $z\sim$~1.3)  based on the GEMS $HST$ survey 
and Chandra Deep Field South data  (Jogee et al. 2004a,b) will  
help constrain  how  bars relate to  AGN activity at these  epochs.

\end{enumerate}

\section{Nuclear Bars}

\subsection{Nuclear Bars: Theory and Observations}

A large-scale bar efficiently drives gas from the outer disk down
to scales of a few hundred pc where the gas inflow stalls after
crossing an ILR. Theory and simulations  (Shlosman et al. 1989; 
Friedli \& Martinet 1993;  Heller \& Shlosman 1994; 
Maciejewski \& Sparke 2000) have suggested  that a nuclear bar (so 
called  `secondary' bar)  nested  within  the large-scale bar (so 
called  `primary' bar)  can break the status quo and gravitationally 
torque digestible fuel closer to the galactic center.
Several scenarios exists for the formation and evolution 
of nuclear bars. The nuclear bar can  decouple from the
primary bar such  that its pattern speed  ($\Omega_{\rm n}$) 
is higher 
(Shlosman et al. 1989; Friedli \& Martinet 1993; Heller \& Shlosman 1994)
or lower 
(Heller, Shlosman \& Englmaier 2001)
than the primary pattern  ($\Omega_{\rm p}$), 
depending on whether it forms via a  
self-gravitational or non-self-gravitational instability.
The nuclear bar can also be in a coupled  phase 
with ($\Omega_{\rm n}$= $\Omega_{\rm p}$),  while 
in the case of a merger remnant  it may even counter-rotate 
with respect to the primary bar.
In simulations, the decoupled phase with $\Omega_{\rm n}$ 
$>$  $\Omega_{\rm p}$  is particularly effective in removing 
angular momentum from the gas and  in helping to fuel the BH.

\begin{figure}
\centering
\hspace{-1.4cm}
%\includegraphics[width=6.3cm]{n2782.figc.ps}
%\hspace{-1.4cm}
%\includegraphics[width=6.3cm]{n2782.figd.ps}
\includegraphics[height=6.3cm]{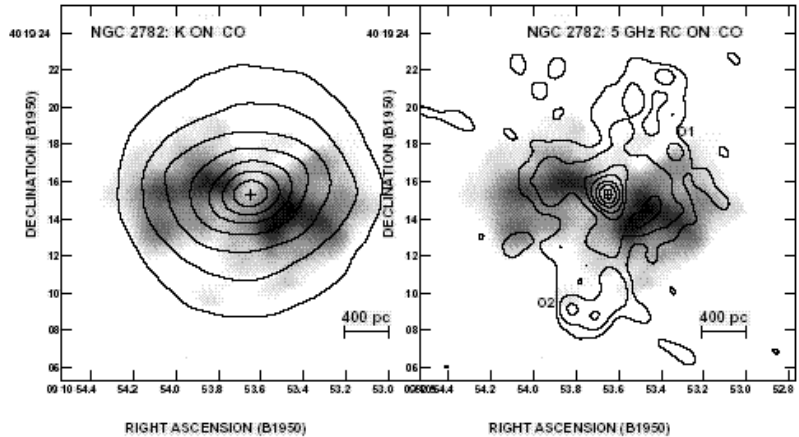}
\caption
{\bf 
A  nuclear stellar bar feeding gas into a powerful starburst
within the inner 100 pc radius of NGC 2782:
\rm
\it
Left: 
\rm
$K$-band contours  are overlaid on the 
($2.1'' \times 1.5''$) CO intensity map  (greyscale).
A nuclear stellar bar (identified via isophotal fitting of the 
$K$-band image) is present at a PA of $\sim$ 100 \deg and is 
itself nested within a large-scale oval/bar which  is visible 
in a larger $I$-band image. The cold gas traced in CO 
has a bar-like distribution which  leads the nuclear 
stellar bar,  and  its velocity field (not shown here) 
reveals weak bar-like streaming motions.
\it
Right:~
\rm
5~GHz radio continuum (contours) are overlaid on the  CO 
map (greyscale).  The nuclear stellar bar   appears to be 
feeding molecular gas into an intense  starburst 
which peaks in RC within the inner 100 pc radius and has
a SFR of 3--6 M$_{\odot}$ yr$^{-1}$.
The starburst in turn appears to be driving an outflow 
associated with the  northern and southern RC 
bubbles. 
%The two CO spurs, labeled O1 and O2, 
%lie inside the  outflow bubbles and are elongated along the 
%CO kinematic axis.
(From Jogee, Kenney, \& Smith 1999)
}
\label{fig:1}       % Give a unique label
\end{figure}

Most observational studies on nuclear bars have focused on the
\it  morphological \rm properties of bars  determined from  
optical and IR  images  (Wozniak et al. 1995; 
Friedli et al. 1996;  Mulchaey \& Regan 1997; 
Jogee, Kenney, \& Smith 1999; 
Laine et al. 2002; Erwin \& Sparke 2002; Erwin 2004).
Studies using space-based and ground-based  IR and optical 
images  reveal that  about 20\%--30\% of S0-Sc galaxies host double 
bars and  20\%--40\% of barred galaxies host a second  bar (e.g., 
Laine et al. 2002; Erwin \& Sparke 2002). 
Confirmed morphologically-identified double bars exist primarily 
in galaxies of Hubble type no later than Sbc. 
They have ellipticities of 0.27--0.6,
semi-major axes of 200--1600 pc, and  position angles which 
both lead and trail the primary bar by varying amounts.
The latter fact suggests that a significant fraction of nuclear
bars must be dynamically decoupled with respect to the primary bar. 

NGC 2782 provides a nice observational case of a nuclear bar driving 
gas down to 100 pc scales. It hosts a nuclear stellar 
bar  which is associated  with $\sim$ $2.5 \times 
10^9$ M$_{\sun}$ of molecular gas and appears to be channeling gas
into the central 100 pc where  an M82-class powerful central starburst
resides  (Fig. 10; Jogee, Kenney, \& Smith 1999).
The nuclear stellar bar is identified via isophotal fits to 
$K$-band image showing a characteristic plateau in position angle 
(PA) as the  
ellipticity rises to a maximum. The molecular gas distribution is
bar-like, leads the nuclear stellar bar, and  shows  weak bar-like 
streaming motions  (Jogee et al. 1999). NGC 2782  may be witnessing 
the  early decoupling  phase of a nuclear stellar bar induced by 
gas inside the OILR of the large-scale bar, which itself appears
to be dissolving.

Aside from NGC 2782,  only a handful of other nuclear bar 
candidates have high resolution CO maps. These show 
a wide range of  CO morphologies  in double-barred galaxies.
The double-barred galaxy NGC 5728 shows CO clumps in a very 
disordered configuration 
(Petitpas \& Wilson 2003). In NGC 4314, the CO emission forms 
a clumpy  ring at the end of the nuclear bar (Benedict et al. 1996; 
Jogee et al. 2004c). The differences in CO  distributions in 
these nuclear bars may be linked to evolutionary differences 
(Jogee et al. 2004c) or differences in gas properties (Petitpas \&
Wilson  2003). 

% Need stellar kinematics
% 

\subsection{Correlations between Nuclear Bars and AGN}

Multiple investigations have searched for a nuclear bar--AGN correlation.
Statistics  from  earlier studies  which used 
offset dust lanes to identify nuclear bars  (e.g., Regan \& Mulchaey 1999; 
Martini \& Pogge 1999) must be re-evaluated because  
% both smoothed particle hydrodynamics SPH and hydrodynamical 
simulations (Maciejewski et al. 2002; 
Shlosman \& Heller 2002) show that  the gas flow 
in  nested nuclear bars differs  fundamentally from that along large-scale 
bars and does not lead to large-scale \it  offset \rm shocks and dust lanes.
% PN: extra reason
% Has time-dependent gravitational potential in ALL frames of reference. 
% Is not associated with  the single-periodic stellar orbits that 
% lead to offset large-scale shocks in primary bars. 
More recent studies based on  isophotal fits  to NIR images 
report that the fraction of   nuclear bars is similar (20\%--30\%) in 
Seyfert and  non-Seyfert hosts (Laine et al. 2002; Erwin \& Sparke 2002). 
There are  several possible explanations for the lack of statistical 
correlations observed between nuclear bars and AGN activity.
\begin{enumerate}
\item
Nuclear bars help to solve the angular momentum problem 
one step further than large-scale bars, but the gas still
has to lose several orders of magnitude in $L$  even at 
pc scales (see Fig. 2).
\item
Not all morphologically-identified nuclear bars are  expected 
to be equally effective in removing angular momentum from the 
gas.  Theoretically, the most effective ones are those with 
$\Omega_{\rm n}$ $>$ $\Omega_{\rm p}$. However, to date 
limited  observations exist on  \it kinematic properties \rm
and  pattern speeds of nuclear bars, and this is an area where
much   progress has yet to be made.
\item
Once large mass concentrations build in the core of the
nuclear bar, it can cause first the nuclear bar, then 
the large-scale primary bar, to dissolve. 
The lifetime of secondary nuclear bars is
not precisely determined, but is expected to be short and 
$\sim$ a few bar rotation timescales. 

\item
As discussed in $\S$ 6.3, only   10$^{6}$ M$_{\odot}$  or 
0.1\%--1\% of the gas present on scales of a few 100 pc  can 
adequately  fuel a Seyfert over a  nominal duty cycle of 
10$^{8}$ years at a rate of 10$^{-2}$ M$_{\odot}$ yr$^{-1}$.
Thus, in lieu of strong fueling mechanisms such as  nuclear bars, 
localized processes  such as SNe shocks and cloud-cloud collisions 
may be adequate for driving one  ambient  10$^{6}$ M$_{\odot}$ 
cloud from 100s of pc down to 10s of pc (see $\S$ 6.3  point 2).

\end{enumerate}

\section{Nuclear Spirals and AGN/Starburst Activity}

\begin{figure}
\centering
\includegraphics [height=4.7in]{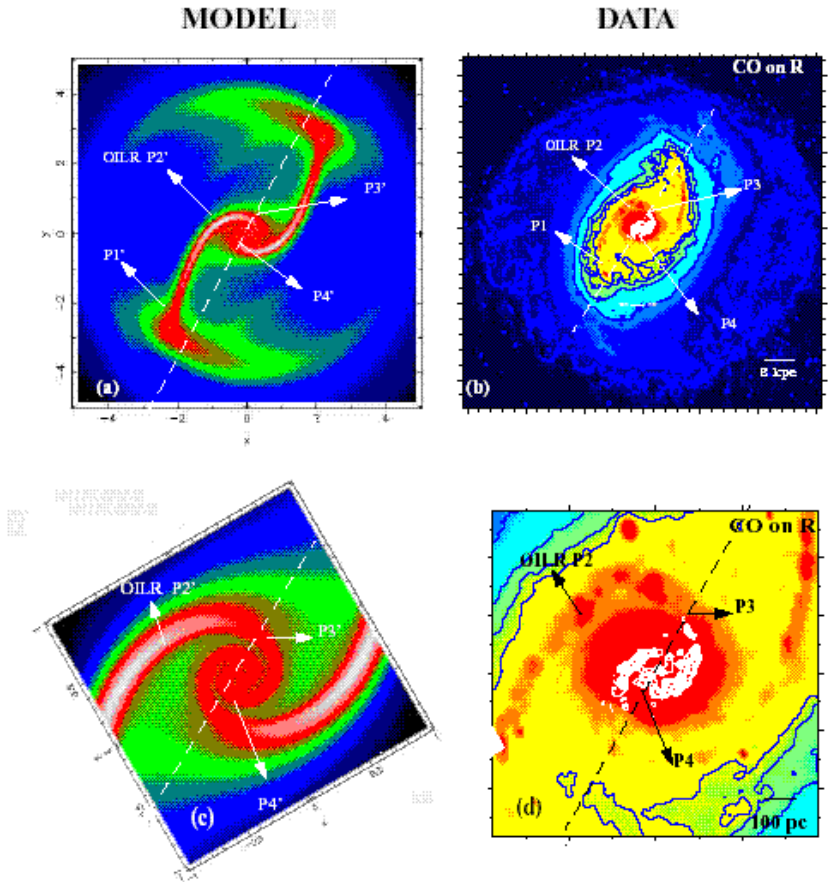}
\caption
{
\bf
Bar-driven gaseous spiral density waves  (SDWs) forming a grand-design 
nuclear dust spiral well inside the OILR of  NGC 5248:
\rm 
Comparison of the  CO and $R$-band data (right) with hydrodynamical  
models   (left; from Englmaier \& Shlosman 2000)  of a bar-driven 
gaseous SDW.  
Points P1' to P4'  in the models  correspond to points P1 to P4 in the data.
The \it top \rm panel shows the entire galaxy, and 
the \it  bottom \rm panel the central  $40 \arcsec$ (3.0~kpc) only.
In NGC 5248, continuous  spirals in CO and dust extend inwards from 6 kpc,  
cross the OILR (P2) of the bar, continue inwards (P3, P4), 
and connect to nuclear dust spirals which extend from 300 pc to 70 pc.
Comparison of the  data with the hydrodynamical  models  
suggest that the continuous connected  
spiral structure  is generated  by  a gaseous  SDW which is 
driven by  the  large-scale bar and  penetrates very deep 
inside the OILR  due to the low central mass concentration
(From Jogee et al. 2002b).
}
\label{fig:1}       % Give a unique label
\end{figure}

\begin{figure}
\hspace{-2.1cm} 
\includegraphics[height=6cm]{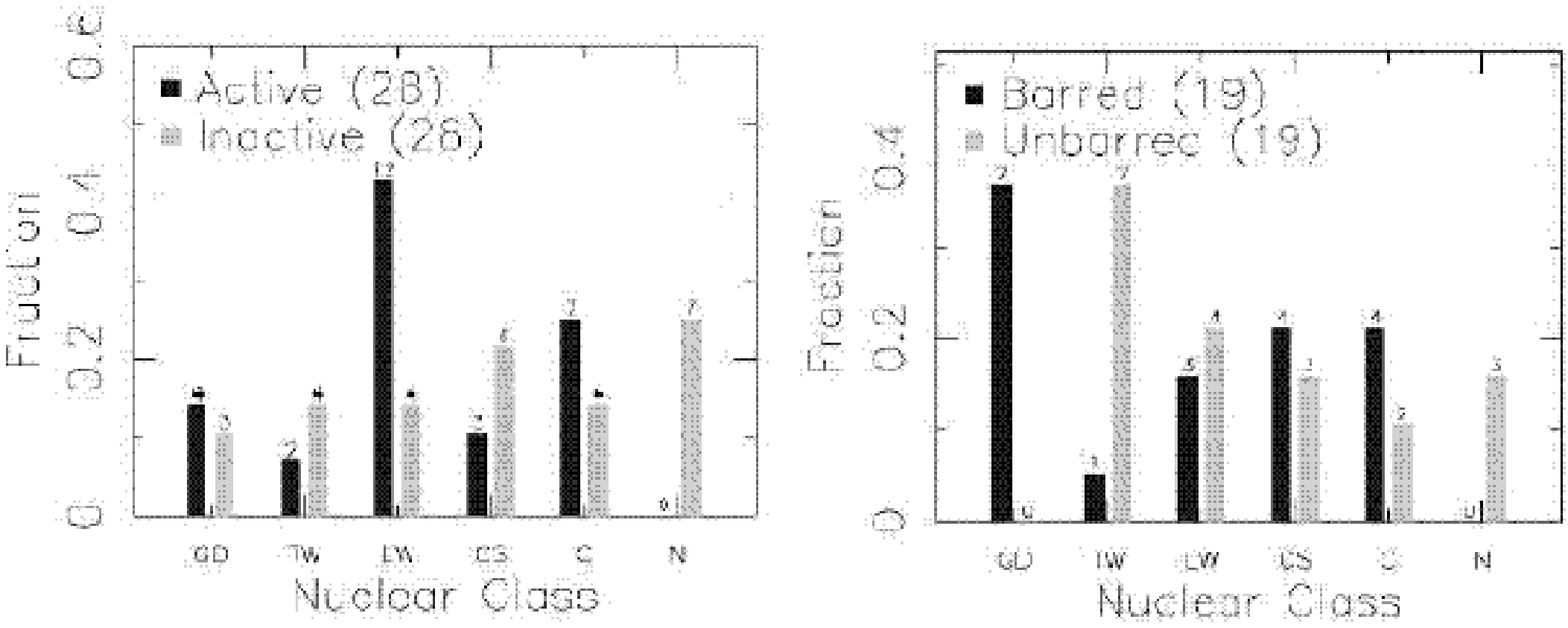}
\caption
{
\bf 
Nuclear dust spirals in AGN hosts and barred galaxies --
\it
Left: 
\rm
Distribution of matched samples of 28 active and 28 inactive galaxies.
The nuclear dust morphology is divided into six classes denoted as 
GD: grand-design nuclear dust spiral, TW: 
tightly wound nuclear spiral, LW: loosely wound nuclear spiral, CS: 
chaotic spiral, C: chaotic dust structure, N: no dust structure present. 
The total number of galaxies in  each class  is  shown in parentheses
and is used to  normalize the histogram bars. 
Nuclear spirals occur with  comparable frequency  
in active and inactive nuclei.
\it
Right: 
\rm
Same as above, but for the matched sample of 19 barred and 19 
unbarred galaxies. 
Grand-design nuclear spirals are found only in barred galaxies.
(Adapted from Martini et al. 2003)
}
\label{fig:1}       % Give a unique label
\end{figure}

There is mounting evidence from  high resolution  
ground-based and \it HST \rm observations  of galaxies 
that nuclear dust spirals are common  on 
scales of a few  tens to hundreds of pc  (e.g.,  
Elmegreen et al. 1998;  Carollo, Stiavelli, \& Mack 1998; 
Laine et al. 1999;  Regan \& Mulchaey 1999; Martini \& Pogge 1999
; Jogee et al. 2002; Martini et al. 2003).
A variety of nuclear spirals are present,  including 
flocculent,  chaotic, and  two-armed grand-design spirals.
Nuclear dust spirals  are likely to  trace shocks since their 
arm-to-interarm  contrast 
(Martini \& Pogge 1999; Elmegreen, Elmegreen \& Eberwein 2002; 
Laine et al. 1999)  implies a mass density  enhancement $\ge$ 2.  
Since shocks dissipate orbital energy and  lead to the outward transfer 
of angular momentum, some authors have suggested  that nuclear dust 
spirals may play some role in the fueling of AGN.

How do nuclear dust spirals  form? 
Most  of them   are believed to be in non-self-gravitating 
central gaseous disks  based on the lack of significant SF observed 
in the spirals. This belief is corroborated in a few cases 
by rough estimates of the neutral gas mass density $\Sigma$(H)  and
Toomre $Q$ values  associated  with the nuclear dust spirals 
(Martini \& Pogge 1999).
% $E$($V$-$H$):  
%   $\Sigma_{\rm H}$ ranges from  one to several tens of 
%  $M_{\tiny \sun}$ pc$^{-2}$  while $Q$ is  typically well above 1. 
In such a  non-self-gravitating gas disk, nuclear spirals may 
form in  several ways.
\bf
(1)
\rm 
Hydrodynamical  simulations  (Englmaier \& Shlosman 2000)
suggest that bar-driven shocks which exist near the ILRs  can trigger gaseous 
SDWs which propagate inwards  across the ILRs, 
and lead to grand-design two-armed nuclear dust spirals.
% In contrast to \it stellar \rm  density waves gaseous density waves 
% triggered by strong  bar shocks can propagate inwards across the ILR(s) 
% according to the density wave theory (e.g., Lin \& Shu 1964; Lin, Yuan, 
% \& Shu 1969; Bertin et al. 1989a,b). 
Observational support for the model comes from detailed  comparisons  
of the simulations with the data in NGC 5248 (Jogee et al.  2002b; 
Fig. 11) which hosts a  grand-design nuclear spiral.
% However, it should be noted that the  gaseous SDW weakens 
% each time it crosses the bar major 
% axis due to a  spray-type divergent  flow (e.g., see the points
% P3' and P4' in Fig. 11c). Thus,  by the time it reaches  
% the central region, it has  a low arm-interarm contrast and  
% mass density, and is unlikely to have strong shocks that 
%  remove angular momentum efficiently.
\bf 
(2)
\rm
Repeated compression from acoustic turbulence followed by 
shearing can lead to flocculent nuclear dust spirals 
(Elmegreen, Elmegreen, \& Eberwein 2002; Montenegro et al. 1999). 
% This mechanism can be important over the central few 100 pc 
% as shearing is quite strong and  the sound speed is comparable 
% to the orbital speed of gas. 
% The origin of this turbulence is 
% unclear  but possibilities include pressure-driven instability 
% (Montenegro et al. 1999);
\bf
(3)
\rm 
Simulations including gas dynamical effects (rotation, shear, shocks), 
thermal cooling effects, and feedback from SF (e.g., Wada \& Norman 2001) 
lead to a multi-phase ISM  where chaotic and flocculent spirals  reccur. 
% Chaotic and flocculent  spirals could be recurrent structures 
% in such a  multi-phase ISM. 
%One caveat though is that SF  in the  simulations  is producing a 
%lot of the shocks that shape the structure of this ISM, and most 
%observed nuclear spirals are not associated with SF.  
%\end{enumerate}
%\end{itemize}

Is there a correlation between nuclear spirals and  AGN activity?
% What is the frequency of nuclear spirals in AGN hosts?
A survey based on  NIR  $HST$ images and color maps of 123 nearby 
galaxies by  Martini \& Pogge (2003) reports  that  nuclear 
dust spirals occur with comparable frequency in both active 
and inactive galaxies (Fig. 12). 
This suggests that the low level inflow rate on scales 
of 10s of pc which might be triggered by shocks in the  
nuclear spirals  is not a universal mechanism for feeding even
low level AGN activity. 
% The only difference they find  
% between the active and inactive galaxies is that
% several inactive galaxies appear to completely lack dust structure 
% in their circumnuclear region, while none of the AGNs lack this 
% structure.
The study also reports that grand-design nuclear 
dust spirals are found only in galaxies  with a large-scale bar,
consistent with afore-mentioned idea that these features 
form via bar-driven  gaseous SDWs (Fig. 11).

\section{From Hundred pc  to Sub-pc Scales}

It is far from understood how gas is driven from scales of 
100 pc  down to sub-pc scales where few  direct observational  
constraints exist.  At 100 pc, matter still needs to reduce 
\it
its specific angular momentum  $L$  
by  a factor of more than  100 
\rm 
before it can reach sub-pc scales, and eventually the last 
stable radius of a massive BH (see Fig. 2). 
I  discuss   a few  mechanisms without attempting to do an 
exhaustive review: 

\begin{description}
\item
$\bullet$
\it 
Dynamical friction and feedback from SF:  
\rm
We already discussed dynamical friction in the context of 
minor mergers where  satellite galaxies located 
at tens of kpc sink  towards  the disk 
of the parent galaxy ($\S$ 5.2) while inducing non-axisymmetric 
instabilities in this disk. 
Since dynamical friction  operates on a timescale which is 
$\propto$ ($R^2$ $v/M$ ln$\Lambda$), it 
becomes increasingly important for massive gas clumps  
%(e.g., 10$^7$--10$^8$  M$_{\odot}$)  
located at small radii.  
%(e.g., a few hundred pc).
For instance, Jogee et al. (1999)  
estimate that t$_{df}$ $\sim$ $ 5 \times 10^{7}$ 
for a  $M \sim$~$10^8$ M$_{\odot}$  gas clump at 
$R \sim$~700 pc 
%where the  rotational speed   
%$V_{\rm c}$~$\sim$~220 km s$^{-1}$ 
using  the approximation below:

\begin{equation}
t_{\rm df} = 7 \times  10^{7} \ {\rm yr}  
\ \left(\frac  {M} {10^{8} M_{\odot}} \right)^{-1}
\ \left(\frac  {V_{\rm c}}{300  \ km \ s^{-1}} \right) 
\ \left(\frac  {R}{700 \ {\rm pc} } \right)^{2}
\end{equation}
 
Feedback effects from powerful circumnuclear starbursts of several 
M$_{\odot}$ yr$^{-1}$ which  are common in the  central few hundred
pc of galaxies (e.g., Jogee et al. 2004c) may also contribute
towards removing angular momentum in localized clouds. 
%The latter may
%provide enough fuel (e.g., $10^6$ M$_{\odot}$) over typical 
%duty cycles (e.g., $10^{8}$ years) to feed low luminosity
%AGN, but  other mechanisms on smaller scales will need to kick in
% to  further  drain out the angular momentum. 

\item
$\bullet$
\it 
Tidal Disruption of Gas Clumps:  
\rm
Numerical investigations by Bekki (2000) suggest that 
once clumps get down to tens of  pc  (for instance 
via dynamical friction),  the tidal gravitational field 
of the MBH transforms the clump into a 
moderately thick gaseous disk or torus.  
A few percent of the gas mass (corresponding to a few 
$\times 10^5$  M$_{\odot}$) can be subsequently transferred 
from this gaseous disk  
to the central sub-parsec  region  around the MBH within  
a few  $ \times 10^6$ yr via viscous torques. 
%Viscous torques  are commonly invoked  on small pc scales  
%(e.g., Shlosman et al. 1989;  Pringle   1996;  Jaffe et al. 1999). 

\item
$\bullet$
\it 
Runaway self-gravitational  instabilities:
\rm
It has been suggested that in a gas-rich nuclear disk,  
self-gravitational  instabilities on repeatedly 
small scales could  lead to three or more 
bars  which are nested within each other 
(e.g., Shlosman et al. 1990). 
However, while the presence of three nested bars or triaxial features 
have been reported in a handful of active galaxies (e.g., 
Friedli et al. 1996; Laine et al. 2002), the current limited spatial 
resolution of NIR surveys  and  molecular gas surveys does not yet  allow  
systematic observational tests of this scenario.

\item
$\bullet$
\it 
Stellar mass loss and disruption of stellar clusters: 
\rm
As pointed out by Ho et al. (1997b), nominal 
stellar mass loss rates 
of $\sim$ $10^{-5}$--$10^{-6}$ M$_{\odot}$ yr$^{-1}$ 
in the central regions of galaxies with  luminosity densities 
of  $10^{3}$--$10^{4}$ M$_{\odot}$ pc$^{-3}$ 
may provide enough fuel for low luminosity AGN.
However, while the mass loss rates may be adequate, we yet
have to identify  mechanisms which can reduce the specific angular 
momentum of the gas down to $10^{24}$ cm$^{2}$~s$^{-1}$  
(see  Fig. 2).
In the same vein, the tidal disruption of 
stellar clusters  (Emsellem \& 
Combes 1997; Bekki 2000)  passing by a  central MBH in 
the accretion disk has been invoked as a source of fuel.   
However, it still remains to be investigated how 
exactly the angular momentum of such clusters would be drained  
and whether this is facilitated by clusters on radial orbits.

\item
$\bullet$
\it 
On pc scales:
\rm
On pc ands sub-pc scales viscous torques 
(e.g., Shlosman et al. 1989;  Pringle   1996),  
warping induced in an accretion disk due to radiation pressure
from a central source (Pringle 1996), 
and hydromagnetic outflows in AGN  (Emmering, Blandford, \& Shlosman  
1992)  have been invoked. 
In the latter model, it is postulated that 
a  hydromagnetic wind containing dense molecular clouds 
is  accelerated radiatively and centrifugally away from an
accretion disk,  removing angular momentum from the disk,  
and forming  the  broad emission lines seen in AGN  
(Emmering et al. 1992).

\end{description}

\section{Summary  and Future Perspectives}

Faced with the fallible task of providing  an `objective'  summary 
of this review,  I can only forewarn the reader with these  perennial 
words:

\begin{description}
\item
\it
\noindent
To command the professors of astronomy to confute their own observations 
is to enjoin an impossibility, for it is to command them to not see what
they do see, and not to understand what they do understand, and to find
what they do not discover. - Galileo Galilei
\rm
\end{description}

\begin{enumerate}
\item
\bf 
Symbiotic evolution of BHs and bulges: 
\rm
The  mass of  a central BH appears to be tightly correlated 
with the stellar velocity dispersion  of the bulge of the host galaxy. 
SMBHs with a wide range of masses  (10$^6$--10$^{10}$ M$_{\odot}$)
follow the same  $M_{\rm bh}$--$\sigma$ relation,  although they
reside in a wide array of host galaxies including 
quiescent early-type (E/S0 Sb--Sc) galaxies, local Seyferts, 
and luminous QSOs out to $z\sim$~3. 
Numerous variants of the $M_{\rm bh}$--$\sigma$  relation
have by now been proposed, including tight correlations between 
$M_{\rm bh}$  and quantities such as the bulge luminosity, the 
light concentration of galaxies, the Sersic index, and 
the mass of the DM halo.
It thus appears that active and quiescent BHs bear a common  
relationship to the surrounding triaxial component of their host 
galaxies over a wide  range of  cosmic epochs and BH masses.
This points towards a symbiotic evolution of BHs  and  the 
central triaxial features of their hosts.

\vspace{0.13cm}
\item 
\bf 
Census and growth epoch of BHs:
\rm
A census of the BH mass density at different epochs suggests 
that accretion with a standard radiative efficiency 
during the quasar era ($z$=0.2--5) can readily account for  the BH 
mass density (few $\times$ 10$^{5}$ M$_{\odot}$   Mpc$^{-3}$)
found in  local ($z<0.1$)  early-type galaxies.
Furthermore, only a small fraction of this  local BH mass 
density is in the form of active Seyfert galaxies, and 
in the latter systems, the inferred  mass accretion rates at the BH 
are typically  10$^{3}$ times lower than in QSOs.
Taken together, the evidence suggests  that 
\it\rm 
there is no significant growth of 
BHs at the present epoch compared to the quasar era. 
\rm 
Therefore, in the context of AGN fueling, 
we should bear in mind that
\it \rm
the dominant fueling mechanisms 
for luminous QSOs out to $z\sim 2.5$  may be markedly
different from those impacting local Seyferts.
\rm 
For instance, tidal interactions and  mergers  
are likely to be more important in activating AGN activity 
at early  epochs than in the present-day.

\vspace{0.13cm}
\item
\it 
\bf 
The angular momentum problem:
\rm
One of the most important challenges in fueling AGN  is 
arguably the angular momentum problem. 
\rm 
The specific angular momentum ($L$) of  matter located at 
a radius of a few kpc must be  reduced by   more than 10$^{5}$  
before it is fit for  consumption by a BH (Fig. 2).
\rm
The angular momentum barrier is a more 
daunting challenge than  the  amount  of gas \it per se \rm.
For instance, while there may be ample material (e.g., 10$^{6}$ M$_{\odot}$ 
clouds) in the inner 200 pc radius to fuel 
typical Seyferts over nominal  duty cycles 
(10$^{8}$ years), the challenge is to understand
what fueling mechanisms can drain their  angular momentum by
99.99\%  so that they are digestible by a BH. 
 
\vspace{0.13cm}
\item 
\bf 
Fueling mechanisms on different scales:  
\rm
\it
\rm 
There is no  universal  fueling mechanism 
which operates efficiently on all spatial scales, from 
several kpc  all the way down to the last stable orbit of a BH.
\rm
Instead,  different fueling mechanisms such as 
gravitational torques (e.g., from large-scale bars and nuclear bars), 
dynamical friction (on massive circumnuclear gas clumps),  
hydrodynamical torques (shocks), and viscous torques 
may relay each other   at different radii  in terms of 
their effectiveness in draining angular momentum. 
According to simulations,
 \it \rm 
large-scale bars  (whether spontaneously or tidally induced) 
are the most efficient mechanisms for driving
large gas inflows from several kpc down to the inner few hundred pc.
\rm  
This holds not only in the case of an isolated barred galaxy, but
also for some classes  of minor (1:10) mergers, most  
intermediate (1:3) mass ratio mergers, and  the early phases 
of most major (1:1) mergers.
During the late stages of a major merger, strongly-varying 
gravitational torques and strong shocks on crossing 
orbits can subsequently  drive the circumnuclear gas further 
in at large rates ($\gg$ 1 M$_{\odot}$ yr$^{-1}$).

\vspace{0.13cm}
\item 
\bf 
Correlations between AGN  and interactions: 
\rm
Statistically significant correlations between 
morphological signs of interactions and AGN 
are seen  in systems with high mass accretion 
rates ($\dot{M}$ $\ge$ 10  M$_{\odot}$ yr$^{-1}$) such as  very 
luminous or radio-loud QSOs.   
The presence of a  correlation  only at very high mass 
accretion rates  holds to reason because such accretion 
rates  are primarily  realized  in nature 
during violent processes  such as 
major/intermediate mass-ratio interactions. 
However, it must be noted that the reverse 
does not hold true:  not \it all \rm  major 
interactions lead to extreme activity because their effectiveness 
in inducing large gas inflows depend on many merger
parameters (speeds, energies, spin-orbital angular momenta alignments).

For moderate  luminosity QSOs and typical  Seyferts,  no clear 
correlation between activity and interactions is seen. 
The lack of a correlation may  be due to 
the fact that  the low $\dot{M}$ required  in  Seyferts and lower  
luminosity QSOs  can be provided not only by strong fueling 
mechanisms  such as interactions which impact the   bulk of the gas, 
but also by  localized low-energetic processes which impact only 
10$^{-3}$--10$^{-2}$ of the circumnuclear gas (see $\S$ 5.4).

\vspace{0.13cm}
\item 
\bf 
Correlations between AGN   and host 
galaxy properties:
\rm

%\begin{enumerate}
\begin{description}
\item
$\bullet$
Local AGN (Sy 1 and Sy 2) and   moderate luminosity
($M_{\rm B}$ $\simeq$ -23)  AGN in the redshift range 
$z$=0.4--1.1  tend to reside primarily in early-type 
(bulge-dominated) galaxies. 
%  In the case of local Seyferts,  the morphological typing 
%  is based on the Hubble Type  (E--Sbc) while for 
%  the intermediate redshift AGN, it is based  either on  
%  single Sersic fits  or on concentration indices.
%Locally H{\sc ii}  nuclei tend to lie in later type 
% systems.

\item
$\bullet$
AGN early-type hosts 
show enhanced blue global rest-frame  optical colors compared 
to early-type inactive galaxies.
This holds in low redshift ($z<0.2$) SDSS
studies, as well as in intermediate redshifts ($z\sim$ 0.5--1.1) 
samples. These colors are  consistent with the presence of young
stellar  populations, suggesting 
that the onset of AGN activity is intimately linked
to the recent onset of global SF in the hosts.
\rm

\item
$\bullet$
The frequency  of large-scale stellar  bars is significant higher
in starburst galaxies than normal galaxies. The bar--starburst 
correlation is consistent with the idea that  a  bar
efficiently drains  angular momentum of gas on exactly the right 
spatial scales (several kpc to a few hundred pc; Fig 2)  
relevant for building the pre-requisite large gas concentrations 
for circumnuclear starbursts.
At this time,  the question of whether Seyferts have an excess of 
large-scale bars compared to inactive galaxies remains open.
Studies based on  high resolution  NIR images, different
samples, and different bar identification methods (ellipse fits,
Fourier methods)  yield different conflicting results.
The reader is referred to $\S$ 6.3 for a detailed discussion.

\item
$\bullet$
Seyferts appear to have 
\it 
weaker
\rm 
bar strengths compared to inactive galaxies. 
These results may lend observational support to the 
long-claimed (but little-tested)  idea that a large build-up 
of mass concentration via gas inflow into the inner 100 pc  
can weaken or even destroy the large-scale bar.   However, the 
question of  whether bars are long-lived or whether they 
dissolve and  reform recurrently  over a Hubble time
is  also highly controversial  at the  moment ($\S$ 6.1).

\item
$\bullet$
The frequency of outer rings and of  (inner + outer) rings 
appears to be  higher by a factor of 3-4  in Seyfert galaxies  
compared to normal galaxies. Various interpretations of this 
strong correlation have been proposed. The rings may be weak 
non-axisymmetric oval distortions or they be remnants of 
dissolved bars. Alternatively, the correlation may be the result 
of both Seyferts and rings existing preferentially in 
early-type systems.

\item
$\bullet$
%Nuclear bars and nuclear dust  spirals  on  scales of few hundreds 
%to few tens of pc are common in galaxies. 
The  frequency  of both  nuclear stellar bars (identified  
morphologically)  and nuclear dust  spirals is found to 
be similar in Seyferts and normal  galaxies. 
This lack of correlation suggests that   these 
features  are not universal mechanisms for fueling an AGN  
or/and  that their lifetime is  short ($\le$ few $\times 10^{8}$ years). 

%\end{enumerate}
\end{description}

\vspace{0.13cm}
\item 
\bf 
Future perspectives:
\rm
I outline below a few of the many exciting developments in AGN research 
we might  wish/expect in the  upcoming decade.
\it 
(a)
\rm 
Now that we have a relatively solid  census of BH mass density 
from $z\sim$~0.1--5 ($\S$ 4), we need to systematically map  
the molecular gas content and structural properties of AGN hosts 
as a function of redshift in order to investigate why 
AGN  activity has declined sharply since  $z\sim$2.5.
NIR integral field spectroscopy on 8m-class telescopes, 
inflowing 24 $\mu$m $Spitzer$ data, and the advent of sub-millimeter  
facilities like the 50-m LMT (\it circa \rm 
2007) and ALMA (\it circa \rm 2010) will provide key constraints. 
\it 
(b)
\rm 
To date, all studies between AGN/starbursts and large-scale bars have 
focused on local galaxies.  Yet, both the cosmic SFR density and AGN 
activity increase  out to t $z \sim$ 1. 
The ongoing work on the impact of bars and interactions 
over the  last 9 Gyr  (out to $z\sim$~1.3) on central 
starbursts and AGN  based on the GEMS $HST$ survey 
and Chandra Deep Field South data  (Jogee et al. 2004a,b) 
will  help constrain  how  bars and external triggers 
relate to the  activities  and  structural evolution of 
galaxies at these  epochs.
\it 
(c)
\rm 
There is a  dire lack of high resolution interferometric
observations of  molecular gas for a \it statistically 
significant sample \rm of  AGN,  even locally
Three ongoing surveys are alleviating this problem.
The Molecular gas in Active and Inactive  Nuclei 
(MAIN; Jogee, Baker, Sakamoto, \& Scoville  2001) 
high resolution, multi-line survey covers forty 
(Seyfert, LINER,  and  H{\sc ii})  nuclei. 
MAIN  aims at constraining the drivers of activity 
levels in galactic nuclei, and 
complements  the ongoing Nuclei of GAlaxies  (NUGA) 
survey  (Garcia-Burillo et al. 2003) and the  multiple line 
survey  of Seyferts (Kohno et al. 2001); 
\it 
d)
\rm 
%To date, the widely used $M_{\rm bh}$--$\sigma$ relation 
%is primarily based on early-type galaxies  (mostly E/SOs and  
%a few Sb--Sbc) with masses in the range of   a few $\times$ 
%(10$^7$--10$^9$) M$_{\odot}$, and on QSOs out to $z\sim$~3 
%($\S$~2.2).  
%It is important to test it 
The advent of  large (30--100 m) 
diffraction-limited telescopes such as the Giant Magellan 
Telescope will help test/extend 
the  $M_{\rm bh}$--$\sigma$ relation for late-type spirals, 
IMBHs,  and  low surface brightness ellipticals.
% To date these
% systems are poorly sampled by this apparently fundamental 
% relation. 
%  \it 
%  (d)
%  \rm 
%  We need kinematic studies focusing on the  stellar kinematics 
%  (sense of rotation and pattern speeds)  of nuclear bars 
%  in order to distinguish  between different formation scenarios 
%  and constrain the stability  of bars; 

\end{enumerate}

\section{Acknowledgements}

For comments and discussions, I thank numerous colleagues, in particular,
L. Ferrarese, R. van der Marel, F. Combes, S. Laine,  
the  GEMS collaboration,  I. Shlosman, N. Grogin, J. H. Knapen, 
\& L. Hernquist. I  acknowledge support from  the National Aeronautics
and Space Administration (NASA) under  LTSA Grant  NAG5-13063
issued through the Office of Space Science.

%
% BibTeX users please use
% \bibliographystyle{}
% \bibliography{}
%
% Non-BibTeX users please follow the syntax
% the syntax of "referenc.tex" for your own citations

%
%%%%%%%%%%%%%%%%%%%%%%%%%%%%%%%%%%%%%%%%%%%%%%%%%%%%%%%%%%%%%%%%%%%%%%  }

%%%%%%%%%%%%%%%%%%%%%%%%%%%%%%%%%%%%%%%%%%%%%%%%%%%%%%%%%%%%%%%%%%%%%%

\printindex
\end{document}